\documentclass[preprintnumbers,aps,prb,twocolumn]{revtex4}

\usepackage{graphicx}
\usepackage{dcolumn} 
\usepackage{amsmath}
\usepackage{amssymb}
\usepackage{bm}      
\usepackage{graphicx}
\usepackage{color}
\usepackage{epsfig,psfrag,subfigure,subeqnarray}
\usepackage{bbm}
\usepackage{enumitem}
\def\lan{\langle}
\def\ran{\rangle}
\def\va{\varepsilon}

\def\vk{{\bf k}}

\def\vq{{\bf q}}
\def\vp{{\bf p}}

\newcommand{\bd}{\begin{equation}}
\newcommand{\ed}{\end{equation}}
\newcommand{\be}{\begin{equation}}
\newcommand{\ee}{\end{equation}}
\newcommand{\bt}{\begin{split}}
\newcommand{\et}{\end{split}}

\newcommand{\bn}{\begin{align}}
\newcommand{\en}{\end{align}}

\newcommand{\bea}{\begin{eqnarray}}
\newcommand{\eea}{\end{eqnarray}}
\newcommand{\ba}{\begin{array}}
\newcommand{\ea}{\end{array}}
\newcommand{\nn}{\nonumber}

\begin{document}
\title {Partition function of $N$ composite bosons}

\author{Shiue-Yuan Shiau$^{1}$, Monique Combescot$^2$  and Yia-Chung Chang$^{3,1}$}
\email{yiachang@gate.sinica.edu.tw}
\affiliation{$^1$ Department of Physics, National Cheng Kung University, Tainan, 701 Taiwan}
\affiliation{$^2$Institut des NanoSciences de Paris, Universit\'e Pierre et Marie Curie, CNRS, 4 place Jussieu, 75005 Paris}
\affiliation{$^3$Research Center for Applied Sciences, Academia Sinica, Taipei, 115 Taiwan}
\date{\today}

\begin{abstract}
The partition function of composite bosons (``cobosons" for short) is calculated in the canonical ensemble, with the Pauli exclusion principle between their fermionic components included in an exact way through the finite temperature many-body formalism for composite quantum particles we recently developed. To physically understand the very compact result we obtain, we first present a diagrammatic approach to the partition function of $N$ elementary bosons. We then show how to extend this approach to cobosons with Pauli blocking and interaction between their fermions. These diagrams provide deep insights on the structure of a coboson condensate, paving the way toward the determination of the critical parameters for their quantum condensation.

\end{abstract}

\maketitle

\section{Introduction}

A century ago, Albert Einstein suggested that as temperature decreases, non-interacting elementary bosons must undergo a phase transition with a macroscopic number of these bosons ``condensed" into the system ground state. Such a condensation occurs below a critical temperature which decreases with the boson number $N$ as $N^{2/3}$. Interest in Bose-Einstein condensation (BEC) has been revived a decade ago by its first experimental realization thanks to advanced cooling and gas trapping techniques\cite{Anderson1995sci,DavisPRL1995,BradleyPRL1995}. These techniques now allow the study of condensation in geometrically different or low-dimensional potential wells in which a fixed number of bosons are trapped. In addition, highly controllable Feshbach resonances\cite{Inouye1998Nat} opened the route to the study of the BEC-BCS crossover in atomic systems\cite{Zwierlein2005Nat}. \

As the effect of interaction between particles decreases with particle density, a condensation similar to the condensation of non-interacting elementary bosons predicted by Einstein should in principle occur in a dilute gas of bosonic particles, i.e., composite particles made of an even number of fermions. And indeed, such a phase transition is now commonly produced in ultra-cold atomic vapors\cite{BECbook}. Yet, Bose-Einstein condensation in the case of semiconductor excitons has been searched for decades$^{7-13}$, even though these particles were for a long time considered as the most promising candidate to evidence this remarkable macroscopic quantum effect: due to their very light effective mass, the exciton quantum degeneracy at density easy to experimentally achieve should occur below a few kelvins while temperatures as low as micro-kelvins are required for atoms. By contrast, evidence of condensation in exciton-polaritons\cite{SnokePhysToday2010} has been demonstrated in semiconductor quantum well embedded inside a microcavity\cite{Deng2002Sci,DengPNAS2003,Deng2006} and more clearly in a trap\cite{Balili2007Sci}.\

One reason for such a long time search could be that, due to their internal degrees of freedom, semiconductor excitons exist in bright and dark states, i.e., excitons coupled or not coupled to light. This coupling goes along with an increase of the bright exciton energy, leaving dark excitons in the lowest-energy state. So, the Bose-Einstein condensate of excitons must be dark, i.e., not coupled to light\cite{MC2007PRL,Roland2012PRL,Dubin2013}. Another reason could be that, in addition to Coulomb interaction between carriers, excitons also interact in a non-standard way through carrier exchanges induced by the Pauli exclusion principle between electrons and between holes. We may wonder if the Pauli exclusion principle at density necessary for condensation does not substantially affect the quantum condensation of a coboson gas. In relation to this question, we wish to mention that, although the BCS wave function ansatz with all Cooper pairs condensed into the same state successfully explains the physical properties observed in conventional superconductors, this Pauli exclusion principle still makes the exact wave function for $N$ Cooper pairs, as deduced from the Richardson-Gaudin procedure, quite different from the BCS wave function ansatz\cite{CrouzeixPRL2011}.

Although quite successful in treating systems of interacting elementary particles, either bosonic and fermionic, conventional many-body formalism is inadequate when it comes to cobosons like the excitons: first, conventional many-body theory such as the Green's function formalism is constructed in the grand canonical ensemble whereas $a_X$-size excitons dissociate through a Mott transition when their number reaches $L^3/a_X^3$, which is the maximum number a sample volume $L^3$ can accommodate. Secondly, conventional many-body theory presumes some kind of Hamiltonian which normally consists of a part for the particle kinetic energy and a part for interaction between particles. But, attempts to construct energy-like effective scatterings between cobosons through a ``bosonization procedure" fail, by nature, to allow exchanges between the particle fermionic components because their fermions must be frozen into a fixed configuration: the problem comes from the fact that fermion exchanges are dimensionless; so, they cannot lead to energy-like scatterings in order to possibly appear in the Hamiltonian. These two reasons led us to seek for a new many-body formalism in which the number of cobosons is fixed.

A zero temperature formalism for composite quantum particles which allows handling fermion exchanges induced by the Pauli exclusion principle in an exact way was proposed by Combescot {\it et al}\cite{moniqPhysRep}. We then extended this coboson formalism to finite temperature\cite{MC2011PRL}, paving the way to solving a large variety of coboson many-body effects. The goal of this work is to derive the partition function in the canonical ensemble based on this finite temperature formalism. Through it, all statistical thermodynamic properties, including the critical temperature for quantum condensation, should be possible to obtain.

To start, we reconsider the partition function of non-interacting elementary bosons. The one commonly known is in the grand canonical ensemble. From it, we can mathematically extract the partition function in the canonical ensemble; in practice, however, its numerical implementation is quite tricky. Here, we instead propose a direct derivation of this canonical partition function based on a recursion relation. Through this recursion relation we are directly led to the well-known compact form for the canonical partition function of non-interacting elementary bosons given in Eq.~(\ref{GenForm:Z_N}). Its diagrammatic representation has the great advantage to allow easy identification of the fully uncondensed, partially condensed and fully condensed contributions. \

To show the power of our diagrammatic approach, next we consider interacting elementary bosons. We show how to perform a many-body expansion of the canonical partition function through a recursion relation similar to the one used for non-interacting bosons. Interestingly, we find that the partition function for interacting elementary bosons maintains the same recursion relation---and the same compact form---as for ideal elementary bosons provided that we add interactions in each $n$-particle entangled configuration. While this is reminiscent of cluster expansion for quantum systems\cite{Kerson1987}, here we do not need to assume the property that the partition functions can be divided into groups of ``connected" particles. They automatically show up.

We then turn to the canonical partition function of $N$ cobosons made of two fermions, like the excitons. After recalling the key commutators of the coboson many-body formalism, we first calculate the recursion relation of this partition function at first order in fermion exchange in the absence of interaction scatterings between these cobosons. Although this can be done through a brute-force use of commutators, we have here chosen to present a physically intuitive way in getting this partition function through the extension of the diagrammatic approach we used for non-interacting elementary bosons. Surprisingly, we find that the coboson partition function can be cast in the same compact form as for non-interacting elementary bosons provided that we take into account the possibility that cobosons exchange their fermionic components due to the indistinguishability in each $n$-particle entangled configuration. Since fermion exchange does not lead to a normal particle-particle potential, this canonical partition function is fundamentally different from the one of interacting elementary bosons previously considered. These diagrams allow us to understand how an elementary boson condensate is affected by fermion exchanges induced by the Pauli exclusion principle.

Then, taking into account interaction between the fermionic components of the cobosons becomes rather straightforward due to similarities between interacting elementary bosons and interacting cobosons, differences coming from additional Pauli exchange processes.

The key result of this work is the recursion relation given in Eq.~(\ref{eq:ZN_CBGf2}) for the canonical partition functions of $N$ cobosons. This recursion relation leads to the partition function in the same compact form as the one of non-interacting elementary bosons. Our result evidences that cobosons do not all condense into the same state, as non-interacting elementary bosons do in a BEC condensate. The similar structure of the elementary boson and coboson partition functions may help us build possible links between condensate wave functions and critical parameters for the BEC's of elementary bosons and excitons. Moreover, the statistical entropy derived from the partition function enables us to study the relation between quantum entanglement in quantum information language and the composite particle bosonic nature\cite{LawPRA2005,Chudzicki2010PRL,MCEPL2011,Tichy2013}.

The present paper is organized as follows: In Sec. \ref{sec:EB}, we briefly introduce the compact form for the canonical partition function of non-interacting elementary bosons. Next we present the diagrammatic approach to derive the recursion relation between canonical partition functions. Then we extend this diagrammatic approach to interacting elementary bosons.
In Sec. \ref{sec:CEAtoN}, we first briefly discuss complexities intrinsic in the coboson systems. We then introduce the interaction expansion which allows us to split the coboson partition function into a non-interacting part and an interacting part. Finally, we use a diagrammatic approach to calculate the partition function at zeroth order and also at first order in interaction scattering with Pauli exchange treated at first order. Consequences and significances of our results are discussed in the end.
\section{Elementary bosons\label{sec:EB}}
\subsection{Ideal(non-interacting) Bose gas\label{ssec:IBG}}
We consider a gas of non-interacting elementary bosons with kinetic energy $\va_\vk=\hbar^2\vk^2/2m$. Since these bosons do not interact, the energy of each $\vk$ state occupied by $N_\vk$ bosons simply is $N_\vk \va_\vk$; so, the partition function for this ideal Bose gas in the canonical ensemble reads, for $\beta=1/k_BT$, as
\be
\bar Z_N^{(0)}=\sum_{\{N_\vk\}_N} e^{-\beta \sum_\vk N_\vk \va_\vk},\label{eq:bZN}
\ee
the sum being taken over all possible boson numbers subject to $\sum_\vk N_\vk=N$.
\subsubsection{Canonical partition function starting from grand canonical ensemble\label{sssec:cpfGCE}}
To lift the constraint in the sum of Eq.~(\ref{eq:bZN}), one commonly turns to the grand partition function with $\mu$ fixed instead of $N$, defined as
\be
\bar Z^{(GC)}=\sum_{N=0}^\infty e^{\beta \mu N}\bar Z_N^{(0)}.\label{eq:Zgc1}
\ee
 A compact form for $\bar Z^{(GC)}$ is easy to obtain by noting that it also reads
\bea
\bar Z^{(GC)}&=&\sum_{N=0}^\infty\sum_{\{N_\vk\}_N} e^{-\beta \sum_\vk N_\vk (\va_\vk-\mu)}\label{eq:Zgc2}\\
&=& \prod_\vk \sum_{N_\vk=0}^\infty e^{-\beta  N_\vk (\va_\vk-\mu)}=\prod_\vk \frac{1}{1-e^{-\beta  (\va_\vk-\mu)}}.\nn
\eea\
The chemical potential $\mu$ is ultimately adjusted for the mean value of the particle number in the grand canonical ensemble to equal the number of bosons at hand.

Equation (\ref{eq:Zgc1}) shows that the partition function in the canonical ensemble, $\bar Z_N^{(0)}$, is just the prefactor of $e^{\beta \mu N}$ in $\bar Z^{(GC)}$. This prefactor can be obtained from the $N^{\rm th}$ derivative of $\bar Z^{(GC)}$ with respect to $e^{\beta \mu}$. It has been shown that this yields a compact form to the canonical partition function which reads as\cite{Ford1971,FeynmanSP}
\be
\bar Z_N^{(0)}=\sum_{\{p_i\}}\!\frac{1}{p_1!}\!\left(\!\frac{z(\beta)}{1}\!\right)^{p_1}\!\frac{1}{p_2!}\!\left(\!\frac{z(2\beta)}{2}\!\right)^{p_2}\!\cdots\!\frac{1}{p_N!}\!\left(\!\frac{z(N\beta)}{N}\!\right)^{p_N}\!\!.\label{GenForm:Z_N}
\ee
The $p_i$'s are a set of non-negative integers such that
\be
N=1p_1+2p_2+\cdots+Np_N,\label{GenForm:Z_Ncon}
\ee
while $z(n\beta)$ is defined as
\be
z(n\beta)=\sum_\vk e^{-n\beta \va_\vk}.\label{eq:def_zb}
\ee
\subsubsection{Direct approach to the canonical partition function\label{sssec:DCPF}}
The above derivation of the canonical partition function, based on derivatives of the partition function in the grand canonical ensemble, is smart but completely formal. It moreover presupposes the knowledge of the partition function in the grand canonical ensemble. We here present a direct derivation of the canonical partition function for a boson number $N$. This derivation is not only useful for possible extension to cobosons, but, through its diagrammatic representation, it provides a physical understanding of the various terms as coming from the fully uncondensed, partially condensed and fully condensed configurations.

Let $|\bar \psi_{\{N_\vk\}_N}\ran$ be normalized $N$-particle eigenstate of the system Hamiltonian $\bar H_0$ with $N_\vk$ bosons having an energy $\va_\vk$. The canonical partition function given in Eq.~(\ref{eq:bZN}) can be rewritten as
\be
\bar Z_N^{(0)}=\sum_{\{N_\vk\}_N} \lan \bar\psi_{\{N_\vk\}_N}|e^{-\beta \bar H_0}|\bar\psi_{\{N_\vk\}_N}\ran.\label{eq:cpf_N}
\ee
We can circumvent the difficulty coming from the restriction, $\sum_\vk N_\vk=N$, in the sum over all possible configurations $\{N_\vk\}_N$ by using the closure relation in the $N$-elementary boson subspace written in terms of single boson operators $\bar B^\dag_\vk$. These operators are such that $(\bar H_0-\va_\vk)\bar B^\dag_\vk|v\ran=0$ where $|v\ran$ denotes the vacuum state, with a commutation relation given by
\be
\left[\bar B_{\vk'},\bar B^\dag_\vk\right]_-=\delta_{\vk'\vk}.\label{eq:EB_com}
\ee
This closure relation reads as
\be
\bar {\rm I}_N=\frac{1}{N!}\sum_{\{\vk\}} \bar B^\dag_{\vk_1}\bar B^\dag_{\vk_2}\cdots \bar B^\dag_{\vk_N}|v\ran\lan v|\bar B_{\vk_N}\cdots \bar B_{\vk_2}\bar B_{\vk_1},\label{eq:INB}
\ee
as can be checked from $\bar {\rm I}_2^2=\bar {\rm I}_2$ and to generalize to $\bar {\rm I}_N$.  Since the $|\bar \psi_{\{N_\vk\}_N}\ran$'s are eigenstates of $\bar H_0$, a closure relation also exists for normalized $|\psi_{\{N_\vk\}_N}\ran$'s, reading as
\be
\bar {\rm I}_N=\sum |\bar\psi_{\{N_\vk\}_N}\ran \lan\bar\psi_{\{N_\vk\}_N}| .\label{eq:INB2}
\ee
By injecting Eq.~(\ref{eq:INB}) in front of $|\bar\psi_{\{N_\vk\}_N}\ran$ in Eq.~(\ref{eq:cpf_N}) and by getting rid of the $|\bar\psi_{\{N_\vk\}_N}\ran$ states through Eq.~(\ref{eq:INB2}), we can rewrite $\bar Z_N^{(0)}$ as
\be
\bar Z_N^{(0)}=\frac{1}{N!}\sum_{\{\vk\}} \lan v|\bar B_{\vk_1}\cdots \bar B_{\vk_N}e^{-\beta \bar H_0}\bar B^\dag_{\vk_N}\cdots \bar B^\dag_{\vk_1}|v\ran.\label{eq:cpf_N2}
\ee
\begin{figure}[t]
\centering
   \includegraphics[trim=5cm 8cm 5cm 6cm,clip,width=3in] {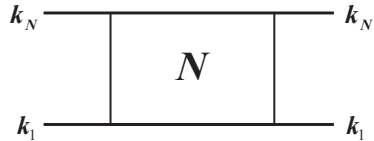}
\caption{\small Scalar product of $N$ elementary bosons appearing in the canonical partition function given in Eq.~(\ref{eq:cpf_N3}).}
\label{fig:1}
\end{figure}
The Hamiltonian $\bar H_0$ for non-interacting elementary bosons reads as $\bar H_0=\sum_\vk \va_\vk \bar B^\dag_\vk \bar B_\vk$; so, the above canonical partition function readily reduces to
\be
\bar Z_N^{(0)}=\frac{1}{N!}\sum_{\{\vk\}} e^{-\beta (\va_{\vk_1}+\cdots +\va_{\vk_N})} \lan v|\bar B_{\vk_1}\cdots \bar B_{\vk_N}\bar B^\dag_{\vk_N}\cdots \bar B^\dag_{\vk_1}|v\ran.\label{eq:cpf_N3}
\ee
Note that (i) the $\vk$'s in the sum now take all possible values without restriction. (ii) a given $\{N_\vk\}_N$ configuration appears once only in Eq.~(\ref{eq:INB2}), while it appears many times in Eq.~(\ref{eq:cpf_N3}), which explains the presence of the $1/N!$ prefactor.
\subsubsection{Recursion relation for $\bar Z_N^{(0)}$\label{sssec:RR} }
The scalar product in the above equation can be calculated using the commutator (\ref{eq:EB_com}). It allows us to replace $\bar B_{\vk_N}\bar B^\dag_{\vk_N}$ by $\delta_{\vk_N\vk_N}+\bar B^\dag_{\vk_N}\bar B_{\vk_N}$. The $\delta_{\vk_N\vk_N}$ term, when inserted into Eq.~(\ref{eq:cpf_N3}), readily gives
\be
\frac{1}{N!}z(\beta)\left[ (N-1)!\bar Z_{N-1}^{(0)}\right].
\ee
To evaluate the  $\bar B^\dag_{\vk_N}\bar B_{\vk_N}$ term, we push the operator $\bar B_{\vk_N}$ to the right according to the commutator (\ref{eq:EB_com}). This yields $(N-1)$ terms like
\be
\delta_{\vk_N\vk_{N-1}}  \lan v|\bar B_{\vk_1}\cdots \bar B_{\vk_{N-1}}\bar B^\dag_{\vk_N}\bar B^\dag_{\vk_{N-2}}\cdots \bar B^\dag_{\vk_1}|v\ran
\ee
which are equivalent when inserted into Eq.~(\ref{eq:cpf_N3}) through a relabeling of the dummy indices $\vk_n$'s. Repeating the same procedure as above, we replace $\bar B_{\vk_{N-1}}\bar B^\dag_{\vk_N}$ by $\delta_{\vk_{N-1}\vk_N}+\bar B^\dag_{\vk_N}\bar B_{\vk_{N-1}}$.
The term in $\delta_{\vk_{N-1}\vk_N}$, when inserted into Eq.~(\ref{eq:cpf_N3}), readily gives
\be
\frac{1}{N!}(N-1)z(2\beta)\left[ (N-2)!\bar Z_{N-2}^{(0)}\right].
\ee
The term in $\bar B^\dag_{\vk_N}\bar B_{\vk_{N-1}}$, calculated by pushing $\bar B_{\vk_{N-1}}$ to the right, yields $(N-2)$ equivalent terms; and so on...

So, we end with a nicely compact recursion relation which simply reads as
\bea
\bar Z_N^{(0)}&=&\frac{1}{N}\left[z(\beta)\bar Z_{N-1}^{(0)}+z(2\beta)\bar Z_{N-2}^{(0)}+\cdots+z(N\beta)\right]\nn\\
&=&\frac{1}{N}\sum_{p=1}^Nz(p\beta)\bar Z_{N-p}^{(0)},\label{eq:rrN}
\eea
with $\bar Z_{0}^{(0)}$ taken as 1. Using this recursion relation, it is easy to recover the expression of the canonical partition function obtained from the grand canonical ensemble\cite{Ford1971}, as given in Eq.~(\ref{GenForm:Z_N}). As illustration, we give the lowest few $\bar Z_N^{(0)}$'s in \ref{app:sec1}.
\begin{figure}[t]
\centering
   \includegraphics[trim=5cm 3.6cm 4cm 4cm,clip,width=2.9in] {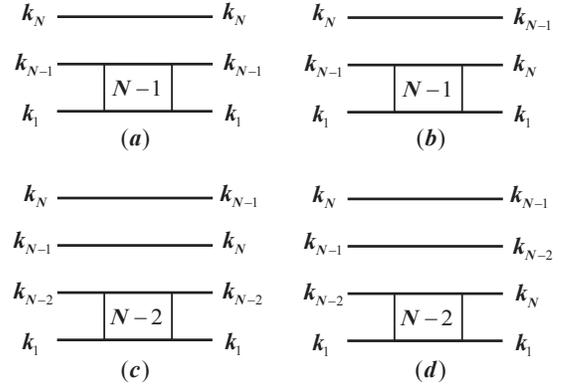}
\caption{\small Diagrams leading to the recursion relation (\ref{eq:rrN}) between the canonical partition functions of non-interacting elementary bosons. }
\label{fig:2}
\end{figure}
\subsubsection{Diagrammatic procedure\label{sssec:DRofCPF}}
It is possible to recover the recursion relation (\ref{eq:rrN}) between the canonical partition functions using diagrams. The diagram of Fig. \ref{fig:1} represents the scalar product of $N$ elementary bosons $(\vk_1,\cdots,\vk_N)$.  We can set up reduction rules to relate this scalar product to those of lower number of bosons. As depicted in Fig.~\ref{fig:2}, this is done by connecting $\vk_N$ on the left to one of the $\vk$'s on the right; this $\vk$ can be either $\vk_N$ as in Fig. \ref{fig:2}(a) (leaving behind a scalar product of $N-1$ bosons) or any other $\vk_n$'s like $\vk_{N-1}$ as in Fig.~\ref{fig:2}(b), which leads to $(N-1)$ similar terms once summation over dummy $\vk$ indices is performed. In the diagram of Fig.~\ref{fig:2}(b), we can connect $\vk_{N-1}$ on the left either to $\vk_N$ as in Fig.~\ref{fig:2}(c) (leaving behind a scalar product of $N-2$ bosons), or to any other $\vk_n$'s like $\vk_{N-2}$ as in Fig.~\ref{fig:2}(d), which leads to $(N-2)$ similar terms once summation over dummy $\vk$'s is performed; and so on...

We then readily find that the process of Fig.~\ref{fig:2}(a) gives to $\bar Z_N^{(0)}$ a contribution equal to $(1/N!) z(\beta)[(N-1)!\bar Z_{N-1}^{(0)}]$. The process of Fig.~\ref{fig:2}(c), which imposes $\vk_N=\vk_{N-1}$, gives a contribution equal to $(N-1)(1/N!) z(2\beta)[(N-2)!\bar Z_{N-2}^{(0)}]$; and so on... So, we do recover the recursion relation between the $\bar Z_N^{(0)}$'s as given in Eq.~(\ref{eq:rrN}), $z(p\beta)$ being the partition function for a condensate made of $p$ elementary bosons, all in the same state.

We are going to show that the partition function for $N$ cobosons obeys a similar recursion relation, provided that we take into account fermion exchanges and interaction scatterings between the composite particles entangled in a condensate. However, before turning to cobosons, let us go one step further by considering interacting elementary bosons. We are going to show that a recursion relation exists provided that we replace $z(n\beta)$ for a non-interacting $n$-boson condensate by a modified $\hat z(n\beta)$ which contains interaction between bosons.
\subsection{Interacting Bose gas\label{ssec:IBG}}
\begin{figure}[t]
\centering
   \includegraphics[trim=4cm 8cm 5cm 6cm,clip,width=3in] {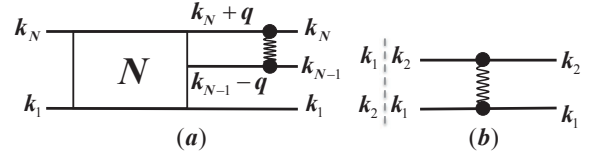}
\caption{\small (a) Diagrammatic representation of $\bar Z_N^{(1)}$. (b) Diagram contributing to $\bar Z_2^{(1)}$ . }
\label{fig:3}
\end{figure}

We now consider interacting elementary bosons. Their Hamiltonian reads
\bea
\bar H&=&\bar H_0+\bar V\label{eq:intEB}\\
 &=&\sum_\vk \va_\vk \bar B^\dag_\vk\bar B_\vk+\frac{1}{2}\sum_{\vk\vk'\vq} V_\vq \bar B^\dag_{\vk+\vq}\bar B^\dag_{\vk'-\vq}\bar B_{\vk'}\bar B_\vk,\nn
\eea
the operators $\bar B^\dag_\vk$ still obeying the commutation relation (\ref{eq:EB_com}). The canonical partition function reads in terms of the $N$-boson eigenstates of the system, $(\bar H-\mathcal{\bar E}_{N,\xi})|\bar \psi_{N,\xi}\ran=0$, as
\be
\bar Z_N=\sum_{\xi}e^{-\beta \mathcal{\bar E}_{N,\xi}}=\sum_\xi \lan \bar \psi_{N,\xi}|e^{-\beta \bar H}|\bar \psi_{N,\xi}\ran.\label{eq:cpf_NInt}
\ee
To get rid of these unknown eigenstates, we follow the same procedure as in Sec. \ref{sssec:DCPF}: we insert the closure relation (\ref{eq:INB}) for $N$ elementary bosons in front of $|\bar \psi_{N,\xi}\ran$ in Eq.~(\ref{eq:cpf_NInt}) and use the fact that $\bar {\rm I}_N=\sum |\bar \psi_{N,\xi}\ran \lan\bar \psi_{N,\xi}|$. The canonical partition function then reads as
\be
\bar Z_N=\frac{1}{N!}\sum_{\{\vk\}} \lan v|\bar B_{\vk_1}\cdots \bar B_{\vk_N}e^{-\beta \bar H}\bar B^\dag_{\vk_N}\cdots \bar B^\dag_{\vk_1}|v\ran.\label{eq:cpf_N2Int}
\ee
Next, we perform a many-body expansion of $\bar Z_N$. We first rewrite $e^{-\beta \bar H}$ using the Cauchy integral formula as
\be
e^{-\beta \bar H}=\int_C \frac{dz}{2\pi i}\frac{e^{-\beta z}}{z-\bar H},\label{eq:ebH_tran}
\ee
where the integration path $C$ is a circle of finite radius centered at the $\bar H$ value on the complex plane. ( For simplicity, we omit this subscript $C$ in the following.) The operator $1/(z-\bar H)$ is expanded for $\bar H=\bar H_0+\bar V$ through
\be
\frac{1}{z-\bar H}=\frac{1}{z-\bar H_0}+\frac{1}{z-\bar H_0}\bar V\frac{1}{z-\bar H} .\label{eq:1overzH}
\ee
This leads us to split the partition function as
\be
\bar Z_N= \bar Z_N^{(0)}+\bar Z_N^{(1)}+\cdots\label{eq:barZNexpan}
\ee
The zeroth-order $\bar Z_N^{(0)}$ in interaction reads as in Eq.~(\ref{eq:cpf_N2}) while the first order is given by
\bea
\lefteqn{\bar Z_N^{(1)}=\frac{1}{N!}\sum_{\{\vk\}} \int \frac{dz}{2\pi i}e^{-\beta z}}\label{eq:cpf_N2Int1st}\\
&&\times\lan v|\bar B_{\vk_1}\cdots \bar B_{\vk_N} \frac{1}{z-\bar H_0}\bar V\frac{1}{z-\bar H_0}\bar B^\dag_{\vk_N}\cdots \bar B^\dag_{\vk_1}|v\ran.\nn
\eea
$\bar H_0$ acting on $\bar B^\dag_{\vk_N}\cdots \bar B^\dag_{\vk_1}|v\ran$ gives $\va_{\vk_1}+\cdots +\va_{\vk_N}$ while
\be
\int \frac{dz}{2\pi i}\frac{e^{-\beta z}}{(z-\bar \va)^2}=-\beta e^{-\beta \va};
\ee
so, $\bar Z_N^{(1)}$ appears as
\bea
\bar Z_N^{(1)}&=&-\beta \frac{1}{N!}\sum_{\{\vk\}} e^{-\beta (\va_{\vk_1}+\cdots +\va_{\vk_N}) }\nn\\
&& \times\lan v|\bar B_{\vk_1}\cdots \bar B_{\vk_N} \bar V\bar B^\dag_{\vk_N}\cdots \bar B^\dag_{\vk_1}|v\ran.\label{eq:cpf_N2Int1st2}
\eea
A convenient way to calculate the above matrix element is to introduce commutators
\bea
\Big[\bar V,\bar B^\dag_\vp\Big]_-&=&\sum_{\vk\vq} V_\vq \bar B^\dag_{\vp+\vq}\bar B^\dag_{\vk-\vq}\bar B_{\vk}=\bar V_\vp^\dag,\label{eq:commut_VB_IBG1}\\
\Big[\bar V_\vp^\dag,\bar B^\dag_{\vp'}\Big]_-&=&\sum_{\vq} V_\vq \bar B^\dag_{\vp+\vq}\bar B^\dag_{\vp'-\vq}.\label{eq:commut_VB_IBG2}
\eea
By pushing $\bar V$ in Eq.~(\ref{eq:cpf_N2Int1st2}) to the right using these commutators, we get $(N-1)+\cdots +1=N(N-1)/2$ terms which contribute equally to $\bar Z_N^{(1)}$ through a relabeling of the dummy indices $\vk_n$'s. By symmetrizing the process, i.e., by also pushing $\bar V$ to the left, we end with the first-order term in interaction reading as
\bea
\lefteqn{\bar Z_N^{(1)}=-\beta \frac{1}{N!}C^N_2\sum_{\{\vk\}} e^{-\beta (\va_{\vk_1}+\cdots +\va_{\vk_N}) }\sum_\vq V_\vq}\label{eq:cpf_N2Int1st3}\\
&&\hspace{-0.4cm} \times\frac{1}{2}\Big[\lan v|\bar B_{\vk_1}\cdots \bar B_{\vk_N} \bar B^\dag_{\vk_N+\vq}\bar B^\dag_{\vk_{N-1}-\vq}\bar B^\dag_{\vk_{N-2}}\cdots \bar B^\dag_{\vk_1}|v\ran+c.c.\Big].\nn
\eea
This matrix element is shown in the diagram of Fig.~\ref{fig:3}(a).

\begin{figure}[t]
\centering
   \includegraphics[trim=0.9cm 2.5cm 2cm 4.5cm,clip,width=3.3in] {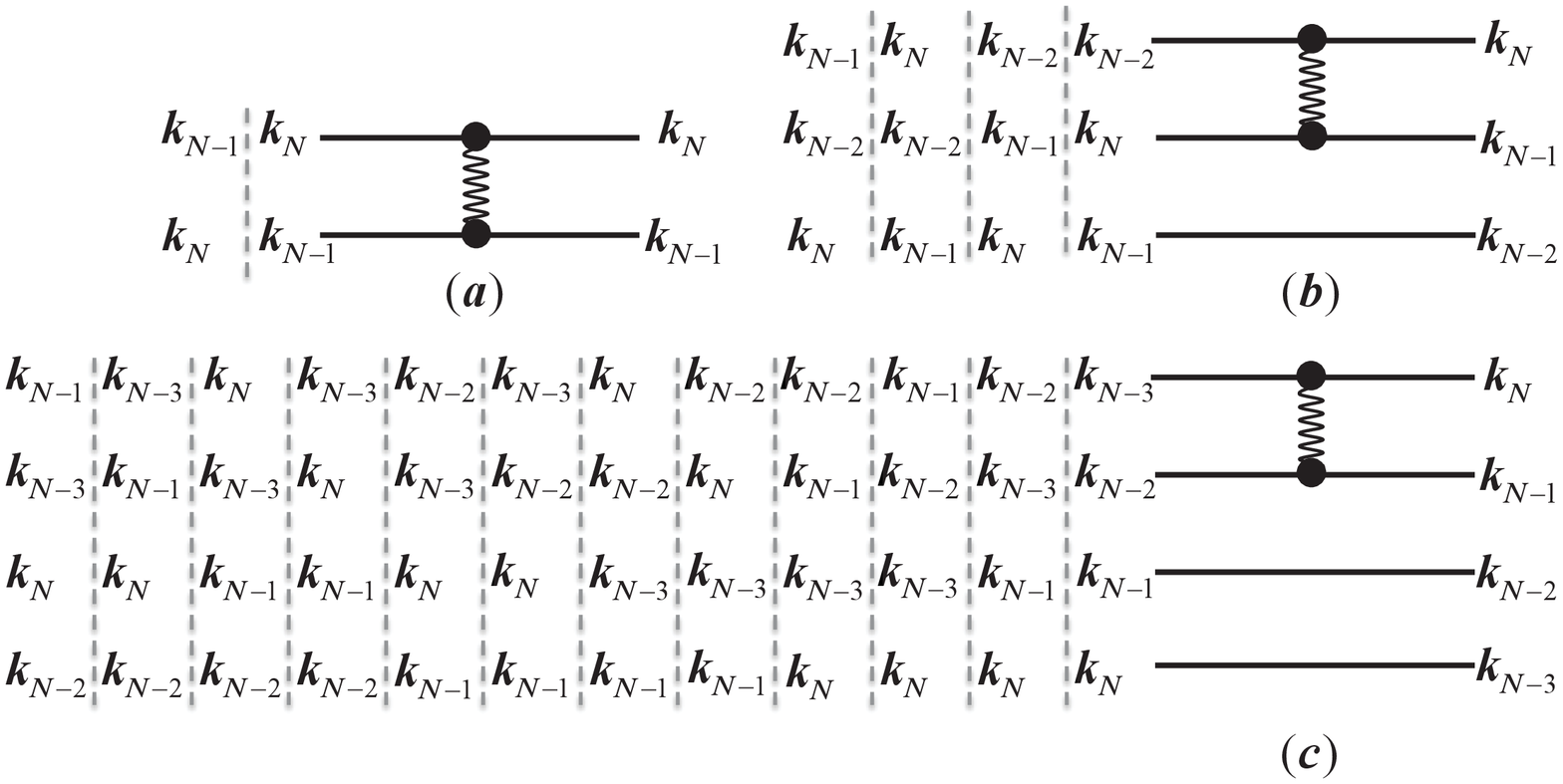}
\caption{\small Three diagrams (a,b,c) corresponding to the prefactors of $\bar Z_{N-2}^{(0)},\bar Z_{N-3}^{(0)}$ and $\bar Z_{N-4}^{(0)}$. }
\label{fig:4}
\end{figure}

For $N=2$, we readily get, since $\bar V \bar B^\dag_{\vk_2}\bar B^\dag_{\vk_1}|v\ran= \sum_\vq V_\vq\bar B^\dag_{\vk_2+\vq}\bar B^\dag_{\vk_1-\vq} |v\ran$,
\be
\bar Z_2^{(1)}=-\frac{\beta}{2} \mathcal{V}(\beta, \beta)
\ee
with $\mathcal{V}(\beta, \beta)$ defined through
\be
\mathcal{V}(n_1\beta, n_2\beta)=\sum_{\vk_1\vk_2} e^{-\beta (n_1\va_{\vk_1} +n_2\va_{\vk_2}) }( V_{\bf 0}+V_{\vk_1-\vk_2}).\label{eq:mathV_IBG}
\ee
$\mathcal{V}(\beta, \beta)$ corresponds to the two processes shown in Fig.~\ref{fig:3}(b), indicated by two columns of $\vk$ vectors $(\vk_1,\vk_2)$ and $(\vk_2,\vk_1)$ separated by a dashed line on the left of the diagram. To understand how the result develops for large $N$, we have explicitly derived $\bar Z_N^{(1)}$ for $N=4$ in \ref{app:sec2}.
For arbitrary $N$, we isolate $\bar Z_{N-2}^{(0)}, \bar Z_{N-3}^{(0)},\cdots$ from the diagram of Fig.~\ref{fig:3}(a) in the same way as for ideal elementary bosons. The prefactor of $\bar Z_{N-2}^{(0)}$ is made of the processes involving $(\vk_N,\vk_{N-1})$ shown in Fig.~\ref{fig:4}(a). Their contribution to $\bar Z_N^{(1)}$ reads as
\be
-\beta \frac{1}{N!}C^N_2 \mathcal{V}(\beta, \beta)\Big[(N-2)! \bar Z_{N-2}^{(0)}\Big]=-\frac{\beta}{2}\mathcal{V}(\beta, \beta)\bar Z_{N-2}^{(0)}.
\ee
The prefactor of $\bar Z_{N-3}^{(0)}$ is made of processes involving $(\vk_N,\vk_{N-1})$ and one of the $(\vk_1,\cdots,\vk_{N-2})$, let say $\vk_{N-2}$. As shown in Fig.~\ref{fig:4}(b) there are four such entangled processes indicated by the four columns of $\vk$ vectors separated by dashed lines on the left of the diagram. In Fig.~\ref{fig:4}(b), $\vk_{N-2}$ ``condenses" either with $\vk_{N-1}$ or with $\vk_N$. Since there are $C_1^{N-2}$ ways to choose this $\vk_{N-2}$ boson among $(\vk_1,\cdots,\vk_{N-2})$, the contribution of such processes to $\bar Z_N^{(1)}$ reads as
\bea
&&-\beta \frac{1}{N!}C^N_2 C_1^{N-2}\Big(\mathcal{V}(\beta,2 \beta)+\mathcal{V}(2\beta, \beta)\Big)\Big[(N-3)! \bar Z_{N-3}^{(0)}\Big]\nn\\
&&=-\frac{\beta}{2}\Big(\mathcal{V}(\beta,2 \beta)+\mathcal{V}(2\beta, \beta)\Big)\bar Z_{N-3}^{(0)}.
\eea
To get the prefactor of $\bar Z_{N-4}^{(0)}$, we isolate two $\vk$'s out of $(\vk_1,\cdots, \vk_{N-2})$, let say $(\vk_{N-2},\vk_{N-3})$. Since there are $C^{N-2}_2$ ways to choose these two $\vk$'s, the contribution to $\bar Z_N^{(1)}$ of the entangled processes between $(\vk_N,\vk_{N-1},\vk_{N-2},\vk_{N-3})$ reads as
 \bea
&&-\beta \frac{1}{N!}C^N_2 C_2^{N-2}\Big(2\mathcal{V}(\beta,3 \beta)+2\mathcal{V}(2\beta, 2\beta)+2\mathcal{V}(3\beta, 1\beta)\Big)\nn\\
&&\times\Big[(N-4)! \bar Z_{N-4}^{(0)}\Big]\nn\\
&&=-\frac{\beta}{2}\Big(\mathcal{V}(\beta,3 \beta)+\mathcal{V}(2\beta, 2\beta)+\mathcal{V}(3\beta, 1\beta)\Big)\bar Z_{N-4}^{(0)}.\label{eq:IBGZN41}
\eea
The three terms in the parentheses originate from the 12 processes shown in Fig.~\ref{fig:4}(c). They correspond to all possible permutations of $(\vk_N,\vk_{N-1},\vk_{N-2},\vk_{N-3})$ on the left which make the same four $\vk$'s on the right entangled, i.e., the $(\vk_{N-2},\vk_{N-3})$ must not ``condense" with themselves; and so on...

 So, we finally get
\be
\bar Z_N^{(1)}=-\frac{\beta}{2}\sum_{n=1}^N \mathcal{\hat V}(n\beta)\bar Z_{N-n}^{(0)}\label{eq:finabarZN1}
\ee
with
\be
\mathcal{\hat V}(n\beta)=\sum_{p=1}^{n-1}\mathcal{V}(p\beta, (n-p)\beta).
\ee
By using the recursion relation between the $\bar Z_N^{(0)}$'s given in Eq.~(\ref{eq:rrN}), we get the partition function of $N$ interacting elementary bosons at first order in interaction as
\be
\bar Z_N\simeq \frac{1}{N}\sum_{n=1}^N\Big[z(n\beta) -\frac{\beta N}{2} \mathcal{\hat V}(n\beta)\Big] \bar Z_{N-n}^{(0)}.\label{eq:ZN_IBGf}
\ee
Note that the second term in the brackets depends on density $N/L^3$ since the $V_\vq$ scattering depends on sample volume as $1/L^3$, which is physically reasonable for many-body effects.

It actually is possible to write $\bar Z_N$ in a compact form like Eq.~(\ref{GenForm:Z_N}). For that, we must transform Eq.~(\ref{eq:ZN_IBGf}) into a recursion relation between the $\bar Z_N$'s similar to Eq.~(\ref{eq:rrN}). To do it, we rewrite $\bar Z_{N-n}^{(0)}$ on the right-hand side of Eq.~(\ref{eq:ZN_IBGf}) in terms of $\bar Z_{N-n}$ using Eq.~(\ref{eq:barZNexpan}). Equation (\ref{eq:ZN_IBGf}) then becomes
\be
\bar Z_N\simeq\frac{1}{N}\sum_{n=1}^N z(n\beta)\Big[\bar Z_{N-n}-\bar Z^{(1)}_{N-n}\big]+\bar Z^{(1)}_N.
\ee
Next, we note that, due to Eq.~(\ref{eq:finabarZN1}),
\be
-\frac{1}{N}\sum_{n=1}^N\! z(n\beta)\bar Z^{(1)}_{N-n} =\frac{1}{N}\frac{\beta}{2}\sum_{n=1}^N\! z(n\beta)\!\!\sum_{m=1}^{N-m}\!\mathcal{\hat V}(m\beta)\bar Z_{N-n-m}^{(0)}.
\ee
As the right-hand side also reads
\bea
\lefteqn{\frac{1}{N}\frac{\beta}{2}\sum_{n=1}^Nz(n\beta)\sum_{m=1}^{N-m}\mathcal{\hat V}(m\beta)\bar Z_{N-n-m}^{(0)}}\nn\\
&=&\frac{1}{N}\frac{\beta}{2}\sum_{n=1}^N\mathcal{\hat V}(n\beta)(N-n)\bar Z_{N-n}^{(0)}\nn\\
&=&-\bar Z^{(1)}_N-\frac{1}{N}\frac{\beta}{2}\sum_{n=1}^Nn\mathcal{\hat V}(n\beta)\bar Z_{N-n}^{(0)},
\eea
we end with $\bar Z_N$ correct up to first order in interaction reading as
\be
\bar Z_N\simeq \frac{1}{N}\sum_{n=1}^N\hat z(n\beta) \bar Z_{N-n}\label{eq:ZN_IBGf2}
\ee
with, for $n\geq 2$,
\be
\hat z(n\beta)=z(n\beta) -\frac{\beta n}{2} \mathcal{\hat V}(n\beta).\label{eq:hatznb}
\ee
It is then straightforward to transform Eq.~(\ref{eq:ZN_IBGf2}) into a compact form like Eq.~(\ref{GenForm:Z_N}) with $z(n\beta)$ replaced by $\hat z(n\beta)$.\

We have demonstrated that, up to first order in interaction, the canonical partition function for interacting elementary bosons takes the same compact form as for non-interacting elementary bosons provided that we replace $z(n\beta)$ by $\hat z(n\beta)$ of Eq.~(\ref{eq:hatznb}). For the perturbative regime to be valid, $N\beta V_{\bf 0}$ must be smaller than 1. Since $V_{\bf 0}$ scales as $1/L^3$, this imposes $N \tilde V/L^3\ll k_BT$. Higher orders in interaction are obtained in the same way using Eq.~(\ref{eq:1overzH}). We then rewrite $\bar Z_{N-n}^{(0)}$'s in terms of $\bar Z_{N-n}$'s to obtain a recursion relation similar to Eq.~(\ref{eq:ZN_IBGf2}).

\section{Composite bosons\label{sec:CEAtoN}}
\subsection{Intrinsic difficulties with cobosons\label{sec:IDComP}}
We now consider cobosons made of two fermions like the excitons. Some difficulties immediately arise when compared to the ideal Bose gas we previously considered.
It is clear that, in order for cobosons to be formed, an attractive interaction between their fermionic components $(\alpha,\beta)$ has to exist. Except for the very peculiar reduced BCS potential in which an up-spin electron with momentum $\vk$ interacts with a down-spin electron with momentum $-\vk$ only, such fermion-fermion interaction automatically brings an interaction between cobosons. \

In addition to this interaction, cobosons also feel each other through the Pauli exclusion principle between their fermionic components. This ``Pauli interaction" in fact dominates most coboson many-body effects. As a result, it is impossible to avoid considering interaction between bosons once we have decided to take into account their composite nature.
\begin{figure}[t]
\centering
   \includegraphics[trim=4cm 3.2cm 4cm 5.5cm,clip,width=3in] {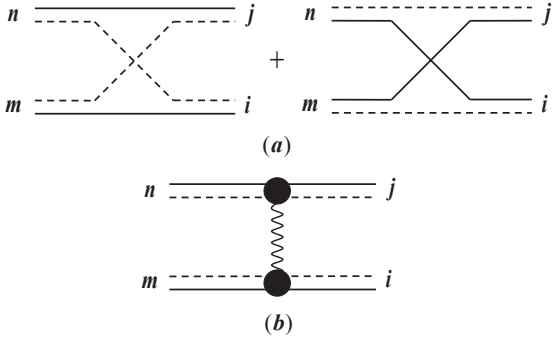}
\caption{\small (a) Pauli scattering $\Lambda\big(^{\hspace{.03cm}
n \hspace{.12cm} j\hspace{.02cm}}_{\hspace{.02cm}
m\hspace{.1cm} i\hspace{.04cm}}\big)$ associated with the exchange of fermion $\alpha$ or $\beta$ in the absence of fermion-fermion interaction. (b) Interaction scattering $\xi\big(^{\hspace{.03cm}
n \hspace{.12cm} j\hspace{.02cm}}_{\hspace{.02cm}
m\hspace{.1cm} i\hspace{.04cm}}\big)$ between the fermions of the cobosons $i$ and $j$, in the absence of fermion exchange. }
\label{fig:5}
\end{figure}
To properly handle many-body effects between cobosons with creation operators
\be
B^\dag_i=\sum_{\vk_\alpha\vk_\beta}a^\dag_{\vk_\alpha}b^\dag_{\vk_\beta}\langle \vk_\beta,\vk_\alpha|i\rangle,\label{Bidag_op}
\ee
where $a^\dag_{\vk_\alpha}$ and $b^\dag_{\vk_\beta}$ are creation operators of their fermionic components, we adopt the commutation formalism introduced in Ref. \onlinecite{moniqPhysRep}:\\
{}(i) Fermion exchanges in the absence of fermion-fermion interaction follow from
\bea
\left[B_m,B^\dag_i\right]_-&=&\delta_{mi}-D_{mi},\label{eq:commut_BB}\\
\left[D_{mi},B^\dag_j\right]_-&=&\sum_n\Lambda\left(\begin{smallmatrix}
n& j\\ m& i\end{smallmatrix}\right)B_n^\dag,\label{eq:commut_DB}
\eea
$D_{mi}$ being such that $D_{mi}|v\ran=0$. The Pauli scattering $\Lambda\left(\begin{smallmatrix}
n& j\\ m& i\end{smallmatrix}\right)$ associated with fermion exchange is shown in Fig.~\ref{fig:5}(a). It corresponds to an exchange of fermion $\alpha$ or $\beta$ between cobosons in states $(i,j)$, which then end in states $(m,n)$. Note that $\Lambda\left(\begin{smallmatrix}
n& j\\ m& i\end{smallmatrix}\right)$ and $\Lambda\left(\begin{smallmatrix}
m& j\\ n& i\end{smallmatrix}\right)$ correspond to the same exchange processes. For simplicity, in the following, we shall use the first diagram with crossing dashed-lines to represent the Pauli scattering $\Lambda\left(\begin{smallmatrix}
n& j\\ m& i\end{smallmatrix}\right)$. \\
\
(ii) Interaction in the absence of fermion exchange follows from
\bea
\left[H,B^\dag_i\right]_-&=&E_iB^\dag_i+V^\dag_i,\label{eq:commut_HB}\\
\left[V^\dag_i,B^\dag_j\right]_-&=&\sum_{mn}\xi\left(\begin{smallmatrix}
n& j\\ m& i\end{smallmatrix}\right)B_m^\dag B_n^\dag,\label{eq:commut_VB}
\eea
$V^\dag_i$ being such that $V^\dag_i|v\ran=0$. The associated interaction scattering $\xi\left(\begin{smallmatrix}
n& j\\ m& i\end{smallmatrix}\right)$ is shown in Fig.~\ref{fig:5}(b).

These four commutators allow us to calculate any many-body effect between cobosons made of fermions $(\alpha,\beta)$, in terms of $\Lambda\left(\begin{smallmatrix}
n& j\\ m& i\end{smallmatrix}\right)$ and $\xi\left(\begin{smallmatrix}
n& j\\ m& i\end{smallmatrix}\right)$, with the Pauli exclusion principle between the fermionic components of these cobosons included in an exact way. The dimensionless parameter which rules many-body effects between $N$ Wannier excitons with Bohr radius $a_X$ in a 3D sample with size $L$, reads as
\be
\eta=N\left(\frac{a_X}{L}\right)^3,\label{def:eta}
\ee
this parameter appearing as $\eta^{n-1}$ in processes in which $n$ excitons are involved.

\subsection{Formal expression of the canonical partition function for cobosons\label{sec:FExCP_ComP}}
The canonical partition function of $N$ cobosons is defined in terms of $N$-pair eigenstate energies, $(H-\mathcal{E}_{N,\xi})|\psi_{N,\xi}\ran=0$, as
\be
Z_N=\sum_\xi e^{-\beta \mathcal{E}_{N,\xi}}=\sum_\xi \lan \psi_{N,\xi}|e^{-\beta H} |\psi_{N,\xi}\ran.\label{eq:ZNC_CP}
\ee
We can get rid of these unknown eigenstates by inserting the closure relation for $N$ cobosons made of two fermions. Instead of Eq.~(\ref{eq:INB}), this closure relation has been shown to read as\cite{MCPRB2008}
\be
{\rm I}_N=\left(\!\frac{1}{N!}\!\right)^2\sum_{\{i\}} B^\dag_{i_1}B^\dag_{i_2}\cdots B^\dag_{i_N}|v\ran\lan v|B_{i_N}\cdots B_{i_2} B_{i_1}.\label{eq:IN_exciton}
\ee
The fact that these cobosons are made of two fermions appears through the prefactor $(1/N!)^2$ instead of $1/N!$.\

By inserting Eq.~(\ref{eq:IN_exciton}) in front of $|\psi_{N,\xi}\ran$ in Eq.~(\ref{eq:ZNC_CP}) and by using the closure relation ${\rm I}_N=\sum_\xi |\psi_{N,\xi}\ran \lan \psi_{N,\xi}| $ for the $N$-pair eigenstates, we can rewrite Eq.~(\ref{eq:ZNC_CP}) as
\be
Z_N=\left(\!\frac{1}{N!}\!\right)^2\sum_{\{i\}} \lan v|B_{i_1}\cdots  B_{i_N} e^{-\beta H} B^\dag_{i_N}\cdots B^\dag_{i_1} |v\ran.\label{eq:ZNC_CP2}
\ee
We wish to stress that difference with the canonical partition function for elementary bosons given in Eq.~(\ref{eq:cpf_N2}) is not so much the prefactor change from $1/N!$ to $(1/N!)^2$ as the fact that the coboson operators $B^\dag_{i}$'s now commute in a  different way from the elementary boson operators. In addition, since these cobosons interact, the Hamiltonian $H$ in $e^{-\beta H}$ cannot be simply replaced by the sum of individual boson energies as in Eq.~(\ref{eq:cpf_N3}).\

To calculate the scalar product of Eq.~(\ref{eq:ZNC_CP2}), we use the commutators for coboson operators given in Eqs.~(\ref{eq:commut_BB}-\ref{eq:commut_VB}). As for interacting elementary bosons, we first use the Cauchy integral formula (\ref{eq:ebH_tran}) to rewrite $e^{-\beta H}$ in order to possibly perform an interaction expansion. This interaction expansion follows from
\be
\frac{1}{z-H}B^\dag_{i}=B^\dag_{i}\frac{1}{z-H-E_i}+\frac{1}{z-H}V^\dag_i\frac{1}{z-H-E_i},\label{eq:int_exp}
\ee
as easy to check using Eq.~(\ref{eq:commut_HB}). So,
\be
e^{-\beta H}B^\dag_i=B^\dag_i e^{-\beta (H+E_i)}+\int \frac{dz }{2\pi i}\frac{e^{-\beta z}}{z-H}V^\dag_i \frac{1}{z-H-E_i}.\label{eq:int_exp2}
\ee
By symmetrizing the expansion procedure, as necessary since we are going to truncate the interaction expansion, as usual in many-body problems, we are led to split $Z_N$ as
\be
Z_N=\sum_{i_N}e^{-\beta E_{i_N}}\big[\Gamma_N(i_N)+I_N(i_N)\big].\label{eq:ZNFI}
\ee
The $I_N(i_N)$ part, which comes from the second term of Eq.~(\ref{eq:int_exp2}), is given by
\bea
\lefteqn{I_N(i_N)=\frac{1}{2}\left(\!\frac{1}{N!}\!\right)^2\!\!\! \sum_{i_1\cdots i_{N-1}}\!\int \frac{dz}{2\pi i} e^{-\beta (z-E_{i_N})}} \label{eq:IN1}\\
&&\hspace{-0.5cm}\times\Big[\lan v| \! B_{i_1}\!\cdots\! B_{i_N}\frac{1}{z-H}V^\dag_{i_N}\frac{1}{z-H-E_{i_N}} B^\dag_{i_{N-1}}\!\cdots \! B^\dag_{i_1}\! |v\ran +c.c.\Big]\nn
\eea
To obtain $I_N(i_N)$ at first order in $V^\dag_i$, we can push the operator $1/(z-H-E_{i_N})$ to the right by only keeping the first term in Eq.~(\ref{eq:int_exp}). This leads to replacing  $H$ on the right of the above matrix element by $E_{i_{N-1}}+\cdots +E_{i_1}$ and $H$ on the left by $E_{i_N}+\cdots +E_{i_1}$. Since
\be
\int \frac{dz }{2\pi i}\frac{e^{-\beta z}}{(z-E_{i_1}-\cdots -E_{i_N})^2}=-\beta e^{-\beta (E_{i_1}+\cdots +E_{i_N})},
\ee
$I_N(i_N)$ appears as
\bea
I_N(i_N)&\simeq&-\beta\left(\!\frac{1}{N!}\!\right)^2 \!\!\! \sum_{i_1\cdots i_{N-1}} \!\!\! e^{-\beta (E_{i_1}+\cdots +E_{i_{N-1}})} \label{eq:IN2}\\
&&\times\frac{1}{2}\Big[\lan v| \! B_{i_1}\!\cdots\! B_{i_N}V^\dag_{i_N} B^\dag_{i_{N-1}}\!\cdots \! B^\dag_{i_1}\! |v\ran+c.c.\Big].\nn
\eea
\begin{figure}[t]
\centering
   \includegraphics[trim=6cm 7.8cm 6cm 7cm,clip,width=2.8in] {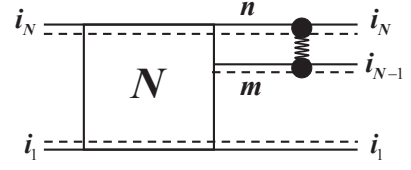}
\caption{\small Scalar product appearing in $\tilde Z_N^{(1)}$ given in Eq.~(\ref{eq:ZNtilde1}).}
\label{fig:6}
\end{figure}
To go further, we use Eq.~(\ref{eq:commut_VB}) to push $V^\dag_{i_N}$ to the right. By noting that $V^\dag_{i_N} B^\dag_{i_{N-1}}\!\cdots \! B^\dag_{i_1}\! |v\ran$ gives $(N-1)$ terms like
\be
\sum_{mn}\xi\left(\!\begin{smallmatrix}
n& i_N\\ m& i_{N-1}\end{smallmatrix}\!\right)B_m^\dag B_n^\dag B^\dag_{i_{N-2}}\!\cdots \! B^\dag_{i_1}\! |v\ran
\ee
which give equal contribution to $I_N(i_N)$ when relabeling the dummy indices $i_n$'s, we end with
\bea
 \lefteqn{I_N(i_N)\simeq-\frac{\beta}{2}\!\left(\!\frac{1}{N!}\!\right)^2\!(N-1)\!\!\sum_{i_1\cdots i_{N-1}}\!\!\! e^{-\beta (E_{i_1}+\cdots +E_{i_{N-1}})}}\label{eq:INiN1}\\
&&\hspace{-0.4cm} \times\Big[\sum_{mn}\lan v| \! B_{i_1}\!\cdots\! B_{i_N} B_m^\dag B_n^\dag B^\dag_{i_{N-2}}\!\cdots \! B^\dag_{i_1}\! |v\ran\xi\left(\!\begin{smallmatrix}
n& i_N\\ m& i_{N-1}\end{smallmatrix}\!\right)+c.c.\Big].\nn
\eea
This term physically corresponds to the diagram of Fig.~\ref{fig:6} in which two out of the $N$ cobosons interact before possibly exchanging their fermions with the other cobosons.

We now consider the $\Gamma(i_N)$ term of $Z_N$ which comes from the first term of Eq.~(\ref{eq:int_exp2}). It reads
\be
\Gamma_N(i_N)=\left(\!\frac{1}{N!}\!\right)^2 \!\!\!\sum_{i_1\cdots i_{N-1}}\!\!\!\lan v| \! B_{i_1}\!\cdots\! B_{i_N}B_{i_N}^\dag e^{-\beta H} B^\dag_{i_{N-1}}\!\cdots \! B^\dag_{i_1}\! |v\ran.\label{eq:ZNgama}
\ee
The same equation (\ref{eq:int_exp2}) leads us to split $\Gamma_N(i_N)$ as
\be
\Gamma_N(i_N)=\sum_{i_{N-1}}e^{-\beta E_{i_{N-1}}}\big[\Gamma_N(i_N,i_{N-1})+I_N(i_N,i_{N-1})\big],
\ee
in which in the second term appears
 \be
\lan v| \! B_{i_1}\!\cdots\! B_{i_N}B_{i_N}^\dag \frac{1}{z-H}V^\dag_{i_{N-1}}\frac{1}{z-H-E_{i_{N-1}}}  B^\dag_{i_{N-2}}\!\cdots \! B^\dag_{i_1}\! |v\ran,
\ee
which is similar to the scalar product appearing in Eq.~(\ref{eq:IN1}) except that we now have $B^\dag_{i_N}$ on the left. Its lowest order in $V^\dag$ is obtained by replacing the right $H$ operator by $E_{i_{N-2}}+\cdots +E_{i_1}$ and the left $H$ operator by $-E_{i_N}+E_{i_{N}}+\cdots +E_{i_1}$. Integration over $z$ in Eq.~(\ref{eq:int_exp2}) again gives $-\beta e^{-\beta (E_{i_1}+\cdots E_{i_{N-1}})}$. So, by symmetrizing the above process, we get
\bea
\lefteqn{I_N(i_N,i_{N-1})\simeq-\beta\left(\!\frac{1}{N!}\!\right)^2 \!\!\! \sum_{i_1\cdots i_{N-2}}\!\!\! e^{-\beta (E_{i_1}+\cdots +E_{i_{N-2}})}}\nn\\
&&\hspace{-0.4cm}\times\frac{1}{2}\Big[\lan v| \! B_{i_1}\!\cdots\! B_{i_N}B^\dag_{i_N} V^\dag_{i_{N-1}}B^\dag_{i_{N-2}}\!\cdots \! B^\dag_{i_1}\! |v\ran+c.c.\Big].
\eea
To go further, we again use Eq.~(\ref{eq:commut_VB}). $V^\dag_{i_{N-1}}B^\dag_{i_{N-2}}\!\cdots \! B^\dag_{i_1}\! |v\ran$ then leads to $(N-2)$ terms similar to
\be
\sum_{mn}\xi\left(\!\begin{smallmatrix}
n& i_{N-1}\\ m& i_{N-2}\end{smallmatrix}\!\right)B_m^\dag B_n^\dag B^\dag_{i_{N-3}}\!\cdots \! B^\dag_{i_1}\! |v\ran
\ee
 which ultimately gives $I_N(i_N,i_{N-1})$ as
\begin{small}
\bea
\lefteqn{I_N(i_N,i_{N-1})\simeq-\frac{\beta}{2}\!\frac{(N-2)}{(N!)^2}\!\!\!\sum_{i_1\cdots i_{N-2}}\!\!\! e^{-\beta (E_{i_1}+\cdots +E_{i_{N-2}})}}  \\
&&\hspace{-0.4cm}\times\Big[\!\sum_{mn}\lan v| \! B_{i_1}\!\cdots\! B_{i_N}B^\dag_{i_N}  B_m^\dag B_n^\dag B^\dag_{i_{N-3}}\!\cdots \! B^\dag_{i_1}\! |v\ran\xi\!\left(\!\begin{smallmatrix}
n&i_{N-1}\\ m& i_{N-2}\end{smallmatrix}\!\right)+c.c.\Big]. \nn
\eea\
\end{small}
To calculate $\Gamma_N(i_N,i_{N-1})$, we proceed in the same way, namely we push $e^{-\beta H}$ in the scalar product to the right using Eq.~(\ref{eq:int_exp2}); and so on...
After summing over $i_N$ and $i_{N-1}$, the $I_N(i_N)$ and $I_N(i_N,i_{N-1})$ terms actually give equal contribution through a relabeling of the $i$'s. So, by considering all equivalent terms, namely $(N-1)+(N-2)+\cdots+1=N(N-1)/2$, we end with
\be
Z_N\simeq Z_N^{(0)}+Z_N^{(1)}\equiv\frac{1}{N!}\big[\tilde Z_N^{(0)}+\tilde Z_N^{(1)}\big],\label{eq:ZNtilde}
\ee
where the zeroth-order term in interaction scattering is
\be
\tilde Z_N^{(0)}=\frac{1}{N!}\sum_{\{i\}}e^{-\beta (E_{i_1}+\cdots E_{i_N})}  \lan v|B_{i_1}\cdots  B_{i_N} B^\dag_{i_N}\cdots B^\dag_{i_1} |v\ran,\label{eq:ZNtilde0}
\ee
while the first-order term in $\xi$ reads as
\begin{small}
\bea
&& \lefteqn{\tilde Z_N^{(1)}=-\frac{\beta}{2} \frac{1}{N!}C^N_2\sum_{\{i\}}e^{-\beta (E_{i_1}+\cdots +E_{i_N})}}  \label{eq:ZNtilde1}\\
&&\times\Big[\!\sum_{mn} \lan v|B_{i_1}\!\cdots  B_{i_N} B_m^\dag B_n^\dag B^\dag_{i_{N-2}}\!\cdots B^\dag_{i_1} |v\ran\xi\!\left(\!\begin{smallmatrix}
n&i_N\\ m& i_{N-1}\end{smallmatrix}\!\right)+c.c.\Big].\nn
\eea
\end{small}
Note that in Eq.~(\ref{eq:ZNtilde}), we have turned from $Z_N$ to $\tilde Z_N$ in order to better see the consequences of the boson composite nature, $\tilde Z_N^{(0)}$ in Eq.~(\ref{eq:ZNtilde0}) and $\bar Z_N^{(0)}$ in Eq.~(\ref{eq:cpf_N3}) then being formally identical: their unique but major difference lies in the commuting relations these $\bar B^\dag_\vk$ and $B^\dag_i$ operators have.

The canonical partition function $Z_N$ in Eq.~(\ref{eq:ZNtilde}) appears as an expansion in interaction scattering $\xi$. In the case of electrons and holes bound into excitons through Coulomb processes, $\xi$ scales as the exciton Rydberg $R_X$ multiplied by the exciton volume $a_X^3$ and divided by the sample volume $L^3$. So, for $N$ excitons, $\tilde Z_N^{(1)}/\tilde Z_N^{(0)}$ scales as $N\beta \xi\simeq \beta R_X \eta $ where $\eta$ is the dimensionless many-body parameter defined in Eq.~(\ref{def:eta}). The many-body interaction expansion we perform is thus valid for $\beta R_X \eta \ll 1$, i.e., $\eta\ll k_BT/R_X$. This ratio is small compared to 1 if the lowest relative motion exciton state only is populated. Note that $\eta$ actively controls the exciton physics because, for $\eta>1$, excitons dissociate into an electron-hole plasma through a Mott transition.
\subsection{Partition function at zeroth order in $\xi$\label{sec:PFComP_0ord}}
To grasp how the Pauli exclusion principle affects the canonical partition function of $N$ cobosons, let us concentrate on its zeroth-order term in interaction scattering given in Eq.~(\ref{eq:ZNtilde0}). The calculation of the scalar product appearing in $\tilde Z_N^{(0)}$ can be done through a brute-force use of Eqs.~(\ref{eq:commut_BB}) and (\ref{eq:commut_DB}). However, as for elementary bosons, calculating this scalar product diagrammatically greatly helps the understanding of the physical processes this part of the partition function contains. This is why here we present a diagrammatic derivation of the recursion relation existing between the $\tilde Z_N^{(0)}$'s, which is similar to the one we gave for elementary bosons. For readers not at ease with diagrams, we also give in \ref{app:sec3} the brute-force calculation of $\tilde Z_N^{(0)}$ for low $N$'s.
\begin{figure}[t]
\centering
   \includegraphics[trim=1.3cm 2cm 2cm 1.8cm,clip,width=3.3in] {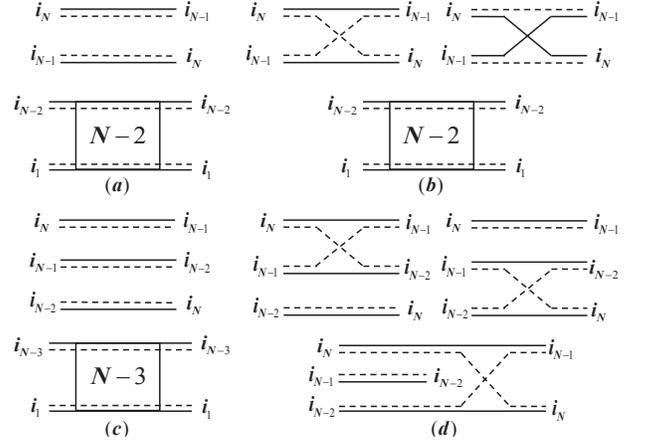}
\caption{\small Diagrams (a,b) correspond to the prefactor of $\tilde Z_{N-2}^{(0)}$ while diagrams (c,d) correspond to the prefactor of $\tilde Z_{N-3}^{(0)}$. }
\label{fig:7}
\end{figure}
\subsubsection{Diagrammatic derivation of recursion relation for $\tilde Z_N^{(0)}$}
The scalar product appearing in $\tilde Z_N^{(0)}$ looks very much like the scalar product of $N$ elementary bosons shown in Fig.~\ref{fig:1}, except that the $\vk_n$ lines are now replaced by $i_n$ double-lines representing the fermions $\alpha_n$ and $\beta_n$ of the coboson $i_n$. As for elementary bosons, we can connect the $i_N$ double-line on the left to the $i_N$ double-line on the right, leaving the $(N-1)$ other cobosons unaffected, in the same way as in Fig.~\ref{fig:2}(a). This process readily leads to a contribution to $\tilde Z_N^{(0)}$ given by
\be
\frac{1}{N!}z(\beta)\left[ (N-1)!\tilde Z_{N-1}^{(0)}\right].
\ee
We can also connect the $i_N$ double-line on the left to one of the $(N-1)$ other double-lines on the right, let say $i_{N-1}$. The $i_{N-1}$ double-line on the left then has to be connected to one of the $i_n$'s on the right; this can be either to $i_N$ or to one of the $(N-2)$ double-lines like $i_{N-2}$. The first process leads to the diagrams (a,b) of Fig.~\ref{fig:7}: since the cobosons $i_N$ and $i_{N-1}$ can exchange their fermions, these two cobosons appear either as in Fig.~\ref{fig:7}(a) or as in Fig.~\ref{fig:7}(b). The physical processes corresponding to these two diagrams bring a contribution to $\tilde Z_N^{(0)}$ given by
\be
(N-1)\frac{1}{N!}\tilde z(2\beta)\left[ (N-2)!\tilde Z_{N-2}^{(0)}\right],
\ee
where $\tilde z(2\beta)=z(2\beta)-L(\beta, \beta)$, the fermion exchange part $L(\beta, \beta)$ being defined through
\be
L(n_1\beta,n_2\beta)=\sum_{i_1 i_2}e^{-\beta(n_1 E_{i_1}+n_2E_{i_2})}\Lambda\left(\begin{smallmatrix}
i_1& i_2\\ i_2& i_1\end{smallmatrix}\right).\label{eq:defaa'l2}
\ee
We now consider processes in which $i_{N-1}$ on the left is connected to $i_{N-2}$ on the right (in the same way as in Fig.~\ref{fig:2}(d)). We can then connect $i_{N-2}$ on the left to $i_N$ or to any of the other $(N-3)$ cobosons like $i_{N-3}$ on the right. The first possibility leads to the diagrams shown in Figs.~\ref{fig:7}(c,d), in which the three cobosons $(i_N,i_{N-1},i_{N-2})$ possibly exchange their fermions. If we restrict to one fermion exchange only, we get the three processes shown in Fig.~\ref{fig:7}(d) in which two cobosons are in the same state, while in the process of Fig.~\ref{fig:7}(c) the three bosons are condensed into the same state. So, the processes of Figs.~\ref{fig:7}(c,d) bring a contribution to $\tilde Z_N^{(0)}$ given by
\be
(N-1)(N-2)\frac{1}{N!}\tilde z(3\beta)\left[ (N-3)!\tilde Z_{N-3}^{(0)}\right],
\ee
where $\tilde z(3\beta)$ at first order in fermion exchange is equal to $z(3\beta)-3L(2\beta, \beta)$.
\begin{figure}[t]
\centering
   \includegraphics[trim=1.9cm 1.2cm 1.7cm 1cm,clip,width=3.3in] {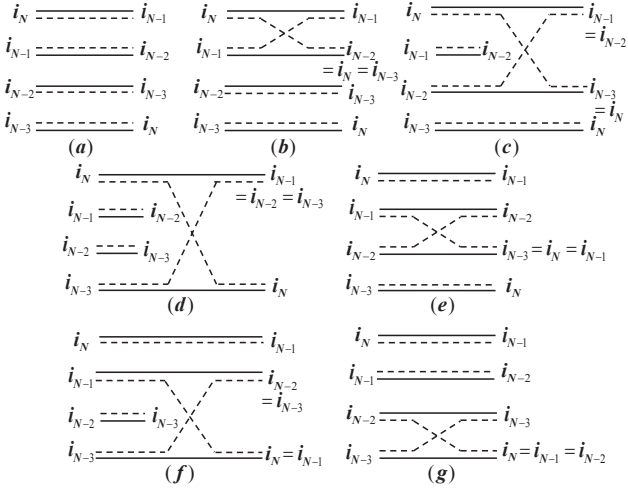}
\caption{\small Diagrams corresponding to the prefactor of $\tilde Z_{N-4}^{(0)}$. }
\label{fig:8}
\end{figure}
To go one step further, we isolate the cobosons $(i_1,\cdots,i_{N-4})$, while the cobosons $(i_N,i_{N-1},i_{N-2},i_{N-3})$ form a condensate in which they possibly exchange their fermions as shown in Fig. \ref{fig:8}. If we restrict to one fermion exchange only, we must connect any two double-lines by exchange, leaving unaffected the other two double-lines, these lines imposing their cobosons to be in the same state. This brings a contribution to $\tilde Z_N^{(0)}$ given by
\be
(N-1)(N-2)(N-3)\frac{1}{N!}\tilde z(4\beta)\left[ (N-4)!\tilde Z_{N-4}^{(0)}\right],
\ee
where $\tilde z(4\beta)$ at first order in fermion exchange is equal to $z(4\beta)-4L(3\beta, \beta)-2L(2\beta, 2\beta)$. The $z(4\beta)$ term comes from diagram (a), the four $L(3\beta, \beta)$ term come from diagrams (b,d,e,g) while the two $L(2\beta, 2\beta)$ term come from diagrams (c,f).

Using the same procedure, we end with the following recursion relation between the $\tilde Z_N^{(0)}$'s
\be
\tilde Z_N^{(0)}=\frac{1}{N}\sum_{n=1}^N \tilde z(n\beta)\tilde Z_{N-n}^{(0)}.\label{eq:tilderrN}
\ee
This is just the one for elementary bosons (\ref{eq:rrN}) but with $z(n\beta)$ replaced by $\tilde z(n\beta)$: $\tilde z(\beta)=z(\beta)$ while, for $n\geq2$, $\tilde z(n\beta)$ reads, at lowest order in fermion exchange,
\be
\tilde z(n\beta)\simeq z(n\beta)-\frac{n}{2}\sum_{m=1}^{n-1}L(m\beta,(n-m)\beta)\label{eq:tildezns}
\ee
with $L(n_1\beta,n_2\beta)=L(n_2\beta,n_1\beta)$, as seen from Eq.~(\ref{eq:defaa'l2}).\

The recursion relation (\ref{eq:tilderrN}) allows us to write $\tilde Z_N^{(0)}$ in the same form as $\bar Z_N$ in Eq.~(\ref{GenForm:Z_N}) with $z(n\beta)$ simply replaced by $\tilde z(n\beta)$. We must however note that, in order to get $\tilde Z_N^{(0)}$ at first order only in fermion exchange, we have to keep one $\tilde z(n\beta)$ only, while taking the other $p$-boson condensates as $z(p\beta)$.
\begin{figure}[t]
\centering
   \includegraphics[trim=5cm 4.3cm 5cm 3cm,clip,width=2.8in] {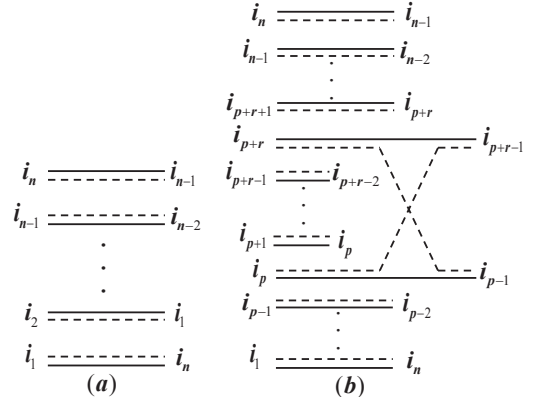}
\caption{\small Diagrams contributing to $\tilde z(n\beta)$. }
\label{fig:9}
\end{figure}
\subsubsection{Partition function of a $n$-coboson condensate at zeroth order in $\xi$}
$\tilde z(n\beta)$ appears as the partition function of a $n$-coboson condensate with fermion exchange between their fermionic components. The diagrammatic representation of the partition function for a $n$-elementary boson condensate is shown in Fig.~\ref{fig:9}(a) with the double-lines replaced by single lines. This diagram indeed imposes $i_n=i_{n-1}=\cdots=i_1$. As these $n$ bosons have the same energy, their partition function is given by $\sum_i e^{-n\beta E_i}=z(n\beta)$. To get the partition function of a $n$-coboson condensate, we must add fermion exchange to this diagram. At first order, this corresponds to processes like the one of Fig.~\ref{fig:9}(b) with one fermion exchange between any two double-lines. The cobosons unaffected by this exchange imposes $i_{p+r-1}=i_{p+r-2}=\cdots=i_p$ and $i_{p-1}=\cdots=i_1=i_n=i_{n-1}=\cdots=i_{p+r}$. So, the diagram (b) brings an exchange term equal to $L(r\beta, (n-r)\beta)$ to the partition function of the $n$-coboson condensate. Due to the various ways $p$ can be chosen and the fact that $L(r\beta, (n-r)\beta)=L((n-r)\beta, r\beta)$, such an exchange leads to a contribution to the partition function of a $n$-coboson condensate given by $(n/2)[L(r\beta, (n-r)\beta)+L((n-r)\beta, r\beta)]$.
Note that, as scatterings involving $n$ cobosons bring a factor $(a_X^3/L^3)^{(n-1)}$, keeping fermion exchange between two cobosons corresponds to performing a many-body expansion at lowest order in density.
\subsection{Partition function at first order in $\xi$}
We now turn to the contribution at first order in interaction scattering to the canonical partition function of $N$ cobosons, as given in Eq.~(\ref{eq:ZNtilde1}) . It is fundamentally similar to the canonical partition function of $N$ interacting elementary bosons given in Eq.~(\ref{eq:cpf_N2Int1st3}). One just has to include fermion exchanges in the processes considered in our previous calculations.

Let us first consider it for $N=2$. It reads
\bea
\tilde Z_2^{(1)}\!\!&=&\!\!-\frac{\beta}{2!}\sum e^{-\beta(E_{i_1}+E_{i_2})}\frac{1}{2}\Big[\lan v|B_{i_1}B_{i_2}B^\dag_m B^\dag_n|v\ran\xi\left(\!\begin{smallmatrix}
m&i_2\\ n& i_1\end{smallmatrix}\!\right)\nn\\
&&+c.c. \Big].\label{eq:Z21CPInt}
\eea
Using the commutators (\ref{eq:commut_BB},\ref{eq:commut_DB}), we find that the scalar product in the above relation reads
as $\delta_{i_1 m}\delta_{i_2 n}+\delta_{i_1 n}\delta_{i_2 m}-\Lambda\left(\!\begin{smallmatrix}
i_1&i_2\\ i_2& i_1\end{smallmatrix}\!\right)$; so, $\tilde Z_2^{(1)}$ is equal to
\be
\tilde Z_2^{(1)}=-\frac{\beta}{2}\hat \xi(\beta, \beta),
\ee
where $\hat \xi(\beta, \beta)$ follows from
\be
\hat \xi(n_1\beta,n_2 \beta)=\sum_{i_1 i_2} e^{-\beta(n_1E_{i_1}+n_2E_{i_2})}\hat \xi(i_1,i_2).\label{def:hatxin1n2}
\ee
The scattering $\hat \xi(i_1,i_2)$ corresponds to all possible direct and exchange interaction processes between incoming cobosons $(i_1,i_2)$ ending in states $(i_1,i_2)$.
It precisely reads
\be
\hat \xi(i_1,i_2)=\xi\left(\!\begin{smallmatrix}
i_2&i_2\\ i_1& i_1\end{smallmatrix}\!\right)+\xi\left(\!\begin{smallmatrix}
i_1&i_2\\ i_2& i_1\end{smallmatrix}\!\right)-\xi^{\rm in}\left(\!\begin{smallmatrix}
i_2&i_2\\ i_1& i_1\end{smallmatrix}\!\right)-\xi^{\rm in}\left(\!\begin{smallmatrix}
i_1&i_2\\ i_2& i_1\end{smallmatrix}\!\right).\label{def:hatxii1i2}
\ee
Precise definition of the exchange scattering $\xi^{\rm in}$ can be found in Ref. \onlinecite{moniqPhysRep}.

$\tilde Z_N^{(1)}$ for arbitrary $N$ is calculated by writing it as a sum of terms proportional to $\tilde Z_{N-p}^{(0)}$. This can be done through a brute-force calculation using the key commutators of the coboson many-body formalism. In \ref{app:sec5}, we show the calculation for $N=3$. Instead, we here give a more enlightening derivation based on diagrams.
\begin{figure}[t]
\centering
   \includegraphics[trim=4cm 3.7cm 4cm 3.5cm,clip,width=3.2in] {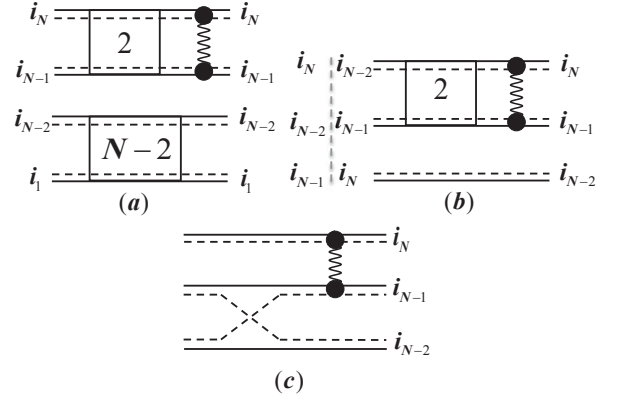}
\caption{\small Interaction processes involving two cobosons (a) and three cobosons (b,c). }
\label{fig:10}
\end{figure}
The scalar product appearing in $\tilde Z_N^{(1)}$ is shown in Fig.~\ref{fig:6}. The prefactor of $\tilde Z_{N-2}^{(0)}$ is made of $(i_N,i_{N-1})$ cobosons only (see Fig.~\ref{fig:10}(a)). It just corresponds to the four direct and exchange interaction processes appearing in $\tilde Z_2^{(1)}$. We readily get their contribution to $\tilde Z_{N}^{(1)}$ as
\be
-\beta \frac{1}{N!} C^N_2 \hat \xi(\beta,\beta)\Big[(N-2)!\tilde Z_{N-2}^{(0)}\Big]=-\frac{\beta}{2}\hat \xi(\beta,\beta)\tilde Z_{N-2}^{(0)}.
\ee
To get the prefactor of $\tilde Z_{N-3}^{(0)}$, we isolate one more cobosons out of $(N-2)$, let say $i_{N-2}$, and we draw all entangled processes. This imposes $i_{N-2}$ not to be connected with itself, as in the diagram of Fig.~\ref{fig:10}(b). By noting that $\hat \xi(\beta,2\beta)=\hat \xi(2\beta,\beta)$, these two processes lead to
\be
-\beta \frac{1}{N!} C^N_2 C^{N-2}_1 \Big[\hat \xi(\beta,2\beta)+\hat \xi(2\beta,\beta)\Big]\Big[(N-3)!\tilde Z_{N-3}^{(0)}\Big].\label{eq:CPZN-31}
\ee
Note that we can also have exchange processes like the ones of Fig.~\ref{fig:10}(c) which connect three cobosons. The associated scatterings, however, are $(a_X^3/L^3)$ smaller than diagram (b). So, the dominant prefactor of $\tilde Z_{N-3}^{(0)}$ is the one given in Eq.~(\ref{eq:CPZN-31}).

As for interacting elementary bosons (see Eq.~(\ref{eq:IBGZN41})), the prefactor of $\tilde Z_{N-4}^{(0)}$ in $\tilde Z_{N-1}^{(1)}$ is obtained by isolating two cobosons out of $(i_1,\cdots, i_{N-2})$, let say $(i_{N-2},i_{N-3})$, and by drawing all entangled processes between $(i_N,i_{N-1},i_{N-2},i_{N-3})$, like in the diagrams of Fig.~\ref{fig:4}(c). This brings a contribution to $\tilde Z_N^{(1)}$ given by
\bea
&&-\beta \frac{1}{N!}C^N_2 C_2^{N-2}\Big[2\hat \xi(\beta,3 \beta)+2\hat \xi(2\beta, 2\beta)+2\hat \xi(3\beta, 1\beta)\Big]\nn\\
&&\times\Big[(N-4)! \tilde Z_{N-4}^{(0)}\Big].\label{eq:CBZN41}
\eea
So, we end with an expansion of $\tilde Z_N^{(1)}$ similar to the one for interacting elementary bosons, $\bar Z_N^{(1)}$, namely
\be
\tilde Z_N^{(1)}=-\frac{\beta}{2}\sum_{n=2}^N \hat \xi(n \beta)\tilde Z_{N-n}^{(0)}\label{def:tildeZN1}
\ee
with
  \be
 \hat \xi(n \beta)=\sum_{p=1}^{n-1}\hat \xi(p \beta,(n-p)\beta). \label{def:hatxinbeta}
  \ee
By adding the Pauli part $\tilde Z_N^{(0)}$ of the $N$-coboson partition function given in Eq.~(\ref{eq:tilderrN}), we find that the canonical partition function of these composite quantum particles is given at lowest order in $(a_X^3/L^3)$ by
\be
\tilde Z_N\simeq \frac{1}{N}\sum_{n=1}^N\Big[\tilde z(n\beta) -\frac{\beta N}{2} \hat \xi(n\beta)\Big] \tilde Z_{N-n}^{(0)}.\label{eq:ZN_CBGf}
\ee
We can go further and transform the above equation into a recursion relation between the $\tilde Z_N$'s by following the procedure we have used for interacting elementary bosons. We then end with $\tilde Z_N$ correct up to first order in both, Pauli exchange and interaction scattering, as
\be
\tilde{Z}_N\simeq \frac{1}{N}\sum_{n=1}^N\Tilde{\Tilde{z}}(n\beta) \tilde{Z}_{N-n}.\label{eq:ZN_CBGf2}
\ee
where the partition function for a $n$-coboson condensate is given, for $n\geq2$, by
\be
\Tilde{\Tilde{z}}(n\beta)=\tilde z(n\beta) -\frac{\beta n}{2} \hat \xi(n\beta).\label{eq:hattildezncb}
\ee
It is then straightforward to show that Eq.~(\ref{eq:ZN_CBGf2}) leads to a compact form for $\tilde{Z}_N$ similar to Eq.~(\ref{GenForm:Z_N}) with $z(n\beta)$ replaced by $\Tilde{\Tilde{z}}(n\beta)$. A similar compact form for the canonical partition function of cobosons to all orders in interaction and fermion exchange appears to us as conceptually obvious, although beyond the scope of the present work.
\section{Conclusions}
We propose a diagrammatic approach to the canonical partition function of $N$ cobosons. In addition to the usual diagrams representing the condensation processes existing for elementary bosons, the Pauli exclusion principle generates new diagrams for fermion exchanges between the fermionic components of cobosons. The partition function we obtain provides grounds for the study of coboson quantum condensation. Here, we calculate in details the canonical partition functions of non-interacting elementary bosons as well as interacting elementary bosons and interacting composite bosons at first order in interaction and fermion exchange. In all cases, the partition function takes the same compact form as the one of non-interacting elementary bosons provided that we include interaction and fermion exchange in the partition function $z(n\beta)$ of the $n$-particle condensate.
\section*{Acknowledgments}
This work is supported by National Cheng-Kung University,  National Science Council of Taiwan under Contract No. NSC 101-2112-M-001-024-MY3, and Academia Sinica, Taiwan. M.C. wishes to thank the National Cheng Kung University and the National Center for Theoretical Sciences (South) for invitations.

\renewcommand{\thesection}{\mbox{Appendix~\Roman{section}}} 
\setcounter{section}{0}
\renewcommand{\theequation}{\mbox{A.\arabic{equation}}} 
\setcounter{equation}{0} %
\section{ $\bar Z_N^{(0)}$ for low $N$'s\label{app:sec1}}
For $N=1$, the canonical partition function reduces to
\be
\bar Z_1^{(0)}=z(\beta).
\ee

For $N=2$, the recursion relation (\ref{eq:rrN}) gives
\be
\bar Z_2^{(0)}=\frac{1}{2!}\left[ z^2(\beta)+z(2\beta)\right]\label{app:Z:N=2}
\ee
in agreement with Eq.~(\ref{GenForm:Z_N}) taken for $(p_1=2)$ or $(p_2=1)$.\

This $\bar Z_2^{(0)}$ taken in the recursion relation for $\bar Z_3^{(0)}$ gives
\be
\bar Z_3^{(0)}=\frac{1}{3!}\left[ z^3(\beta)+3z(\beta)z(2\beta)+2z(3\beta)\right],\label{Z:N=3}
\ee
which agrees with Eq.~(\ref{GenForm:Z_N}) taken for $(p_1=3)$, $(p_1=1,p_2=1)$ or $(p_3=1)$.\

These $\bar Z_1^{(0)},\bar Z_2^{(0)}$ and $\bar Z_3^{(0)}$ taken in the recursion relation (\ref{eq:rrN}) for $N=4$ give
\bea
\bar Z_4^{(0)}&=&\frac{1}{4!}\left[ z^4(\beta)+6z^2(\beta)z(2\beta)+8z(\beta)z(3\beta)\right.\nn\\
&&\left.+6 z(4\beta)+3z^2(2\beta)\right]\label{Z:N=4}
\eea
in agreement with Eq.~(\ref{GenForm:Z_N}) taken for $(p_1=4)$, $(p_1=2,p_2=1)$, $(p_1=p_3=1)$, $(p_4=1)$ or $(p_2=2)$. We note that the sum of prefactors in these partition functions, e.g., $(1+6+8+6+3)/4!$ in the case of 4 bosons, is equal to 1. So, these prefactors physically correspond to the probability of the condensation process at hand.

\renewcommand{\theequation}{\mbox{B.\arabic{equation}}} 
\setcounter{equation}{0} %
\section{Calculation of $ \bar Z_4^{(1)}$\label{app:sec2}}
\begin{figure}[t]
\centering
   \includegraphics[trim=5cm 5cm 5cm 6cm,clip,width=3in] {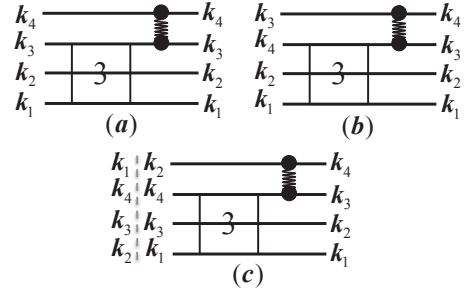}
\caption{\small Diagrams appearing in the scalar product for 4 elementary bosons at first order in interaction (wavy lines). }
\label{fig:11}
\end{figure}
\begin{figure}[t]
\centering
   \includegraphics[trim=2cm 5.5cm 2cm 5cm,clip,width=3in] {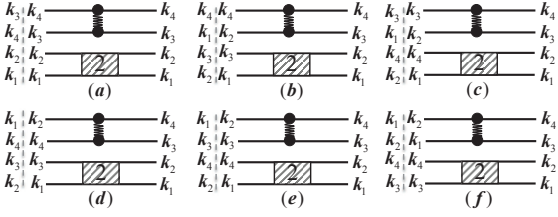}
\caption{\small Diagrams following from the diagrams of Fig.~\ref{fig:11} after we have chosen to connect $\vk_3-\vq$ to one of the three $\vk$'s on the left. }
\label{fig:12}
\end{figure}
\begin{figure}[t]
\centering
   \includegraphics[trim=4.5cm 5.5cm 5.5cm 5cm,clip,width=3in] {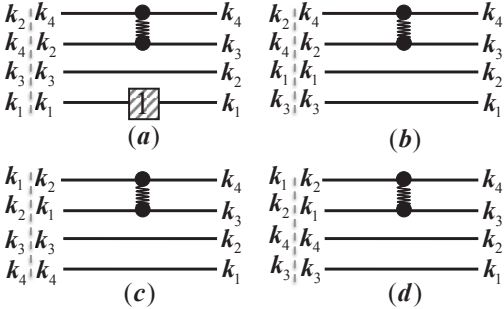}
\caption{\small Diagrams (a,b) follow from the diagrams of Figs.~\ref{fig:12}(b,d), while diagrams (c,d) follow from the diagram of Fig.~\ref{fig:12}(f).}
\label{fig:13}
\end{figure}
The interaction part of the partition function for 4 interacting elementary bosons appears as
\bea
\lefteqn{\bar Z_4^{(1)}=-\beta\frac{1}{4!}C^4_2\sum e^{-\beta(\va_{\vk_1}+\cdots+\va_{\vk_4})}\sum_\vq V_\vq}\\
&&\hspace{-0.3cm}\times \frac{1}{2}\Big[\lan v|\bar B_{\vk_1}\bar B_{\vk_2}\bar B_{\vk_3}\bar B_{\vk_4}\bar B^\dag_{\vk_4+\vq} \bar B^\dag_{\vk_3-\vq}\bar B^\dag_{\vk_2}\bar B^\dag_{\vk_1}|v\ran+c.c.\Big].\nn
\eea
The above scalar product is shown in Fig.~\ref{fig:3}(a) taken for $N=4$. To get it, we can connect $\vk_4+\vq$ to any of the $(\vk_1,\vk_2,\vk_3,\vk_4)$ on the left, as shown in Fig.~\ref{fig:11}. Since connecting $\vk_4+\vq$ to $\vk_2$ or to $\vk_1$ is equivalent, the processes of diagram \ref{fig:11}(c) are going to appear twice.\\
{}
(i) To start, we can connect $\vk_3-\vq$ to $\vk_3$ in diagram \ref{fig:11}(a), and we can connect $\vk_3-\vq$ to $\vk_4$ in diagram \ref{fig:11}(b). These two processes lead to the diagrams shown in Fig.~\ref{fig:12}(a). Their contribution to $\bar Z_4^{(1)}$ reads as
\be
-\beta \frac{1}{4!}C^4_2 \mathcal{V}(\beta,\beta)\Big[2!\bar Z_2^{(0)}\Big]=-\frac{\beta}{2} \mathcal{V}(\beta,\beta)\bar Z_2^{(0)}.
\ee
In diagrams \ref{fig:11}(a) or (b), we can also connect $\vk_3-\vq$ to $\vk_2$ or to $\vk_1$, which gives equivalent contribution; so, these processes, shown in Figs.~\ref{fig:12}(b,c), will appear twice.

Finally, from diagram \ref{fig:11}(c), we can connect $\vk_3-\vq$ to $\vk_4, \vk_3$ or $\vk_1$, as shown in Figs.~\ref{fig:12}(d,e,f).\\
{}
(ii) To go further, we consider diagrams \ref{fig:12}(b,d), and we connect $\vk_2$ to $\vk_3$ or to $\vk_1$, as shown in Figs.~\ref{fig:13}(a,b). Diagram \ref{fig:13}(a) gives a contribution to $\bar Z_4^{(1)}$ equal to
\be
-2\beta \frac{1}{4!}C^4_2 \mathcal{V}(\beta,2\beta)\bar Z_1^{(0)}=-\frac{\beta}{2} \mathcal{V}(\beta,2\beta)\bar Z_1^{(0)},
\ee
while diagram \ref{fig:13}(b) gives a contribution to $\bar Z_4^{(1)}$ equal to
\be
-2\beta \frac{1}{4!}C^4_2 \mathcal{V}(\beta,3\beta)=-\frac{\beta}{2} \mathcal{V}(\beta,3\beta).
\ee

If we now consider diagram \ref{fig:12}(e), we note that it follows from diagram \ref{fig:12}(b) by interchanging $\vk_3$ and $\vk_4$. This interchange also transforms diagram \ref{fig:12}(c) into diagram \ref{fig:12}(d). So, diagrams \ref{fig:12}(c) and (e) give the same contribution as diagrams \ref{fig:12}(d) and (b).

 Finally, in diagram \ref{fig:12}(f) we can connect $\vk_2$ to $\vk_3$ or to $\vk_4$ as shown in Figs.~\ref{fig:13}(c,d). This brings a contribution to $\bar Z_4^{(1)}$ given by
\be
-2\beta \frac{1}{4!}C^4_2 \mathcal{V}(2\beta,2\beta)=-\frac{\beta}{2} \mathcal{V}(2\beta,2\beta).
\ee
Collecting all the terms and noting that $\mathcal{V}(n_1\beta,n_2\beta)=\mathcal{V}(n_2\beta,n_1\beta)$, we end with
\bea
\bar Z_4^{(1)}&=&-\frac{\beta}{2}\Big\{\mathcal{V}(\beta,\beta)\bar Z_2^{(0)} + \big[\mathcal{V}(\beta,2\beta)+\mathcal{V}(2\beta,\beta)\big]\bar Z_1^{(0)}\nn\\
&&+\mathcal{V}(\beta,3\beta)+\mathcal{V}(2\beta,2\beta)+\mathcal{V}(3\beta,\beta)\Big\}.
\eea
\begin{figure}[t]
\centering
   \includegraphics[trim=5cm 4.2cm 5cm 4cm,clip,width=2.8in] {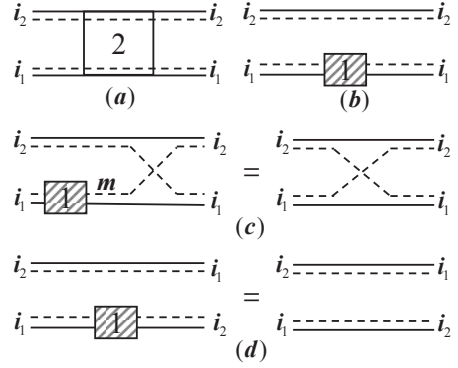}
\caption{\small Diagram (a) represents the scalar product of two cobosons. Diagrams (b,c,d) show the three distinct configurations for these two cobosons.  }
\label{fig:14}
\end{figure}

\renewcommand{\theequation}{\mbox{C.\arabic{equation}}} 
\setcounter{equation}{0} %
\section{Direct calculation of $\tilde Z_N^{(0)}$\label{app:sec3}}

We here show how to calculate the canonical partition function of $N$ cobosons at zeroth order in interaction scattering by using the key commutators (\ref{eq:commut_BB}) and (\ref{eq:commut_DB}) of the many-body formalism. This part of the partition function reads as $Z_N^{(0)}=\tilde Z_N^{(0)}/N!$ with
\be
\tilde Z_N^{(0)}=\frac{1}{N!}\sum_{\{i\}}e^{-\beta (E_{i_1}+\cdots E_{i_N})}  \lan v|B_{i_1}\cdots  B_{i_N} B^\dag_{i_N}\cdots B^\dag_{i_1} |v\ran.\label{app:eq:ZNtilde0}
\ee
To understand how the recursion relation for the $\tilde Z_N^{(0)}$'s given in Eq.~(\ref{eq:tilderrN}) develops, let us explicitly calculate $\tilde Z_N^{(0)}$ for $N=2$ and $N=3$.
\subsection{Two cobosons}
Equation (\ref{eq:commut_BB}) allows us to write the scalar product of two cobosons shown in Fig.~\ref{fig:14}(a) as
\be
\lan v| B_{i_1}B_{i_2}B_{i_2}^\dag B_{i_1}^\dag|v\ran=\lan v| B_{i_1}(\delta_{i_2 i_2}-D_{i_2 i_2}+B_{i_2}^\dag B_{i_2}) B_{i_1}^\dag|v\ran
\ee
By inserting the term in $\delta_{i_2 i_2}$ into $\tilde Z_2^{(0)}$, we readily get its contribution to $\tilde Z_2^{(0)}$ as
\be
\frac{1}{2!}z(\beta)\tilde Z_1^{(0)}.
\ee
The corresponding diagram is shown in Fig.~\ref{fig:14}(b).

Using Eq.~(\ref{eq:commut_DB}) for the term in $D_{i_2 i_2}$, we get
\be
-\sum_m \lan v| B_{i_1} B_m^\dag|v\ran \Lambda\left(\!\begin{smallmatrix}
m& i_1\\ i_2& i_2\end{smallmatrix}\!\right)=-\Lambda\left(\!\begin{smallmatrix}
i_1& i_1\\ i_2& i_2\end{smallmatrix}\!\right).
\ee
The corresponding diagram is shown in Fig.~\ref{fig:14}(c). When inserted into $\tilde Z_2^{(0)}$, this term leads to $-L(\beta, \beta)$. \

Finally, the term in $B_{i_2}^\dag B_{i_2}$ gives $\lan v| B_{i_1} B_{i_2}^\dag|v\ran \delta_{i_1i_2}$ as shown in Fig.~\ref{fig:14}(d). This imposes $i_1=i_2$ and yields $z(2\beta)$. So, we end with
\be
\tilde Z_2^{(0)}=\frac{1}{2!}\Big[z(\beta)\tilde Z_1^{(0)}+(z(2\beta)-L(\beta, \beta))\Big]
\ee
with $\tilde Z_1^{(0)}=z(\beta)$. We can rewrite this expression as Eq.~(\ref{eq:tilderrN}), with $\tilde z(2\beta)=z(2\beta)-L(\beta, \beta)$ in agreement with Eq.~(\ref{eq:tildezns}) taken for $N=2$.

\begin{figure}[t]
\centering
   \includegraphics[trim=5cm 4.7cm 5cm 4.5cm,clip,width=2.8in] {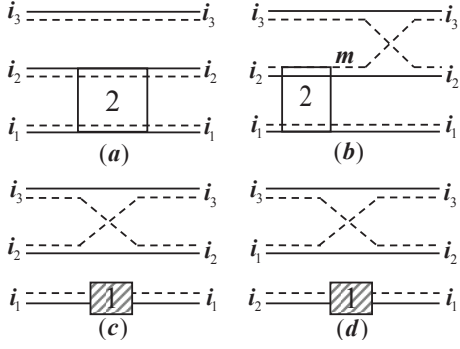}
\caption{\small Diagrams (a,b) representing the first two terms of the three-coboson scalar product in Eq.~(\ref{eq:app_sp3}). Diagrams (c,d) follow from diagram (a). }
\label{fig:15}
\end{figure}
\subsection{Three cobosons}
Equation (\ref{eq:commut_BB}) gives the scalar product of three cobosons as
\bea
\lefteqn{\lan v| B_{i_1}B_{i_2}B_{i_3} B_{i_3}^\dag B_{i_2}^\dag B_{i_1}^\dag|v\ran}\label{eq:app_sp3}\\
&&=\lan v| B_{i_1}B_{i_2}(\delta_{i_3 i_3}-D_{i_3 i_3}+B_{i_3}^\dag B_{i_3}) B_{i_2}^\dag B_{i_1}^\dag|v\ran.\nn
\eea
The term in $\delta_{i_3 i_3}$, when inserted into Eq.~(\ref{app:eq:ZNtilde0}) taken for $N=3$, readily yields a contribution to $\tilde Z_3^{(0)}$ given by
\be
\frac{1}{3!}z(\beta)\Big[2!\tilde Z_2^{(0)}\Big],
\ee
which corresponds to the diagram of Fig.~\ref{fig:15}(a).

For the term in $D_{i_3 i_3}$ of Eq.~(\ref{eq:app_sp3}), we use Eq.~(\ref{eq:commut_DB}) to replace $D_{i_3 i_3}B_{i_2}^\dag$ by $\sum_m B^\dag_m \Lambda\left(\!\begin{smallmatrix}
m& i_2\\ i_3& i_3\end{smallmatrix}\!\right)+B_{i_2}^\dag D_{i_3 i_3}$ and we use again Eq.~(\ref{eq:commut_DB}) for $D_{i_3 i_3}B_{i_1}^\dag$. This leads to
\be
-\sum_m \lan v| B_{i_1}B_{i_2} B_m^\dag \Big[\Lambda\left(\!\begin{smallmatrix}
m& i_2\\i_3& i_3 \end{smallmatrix}\!\right)B^\dag_{i_1}+\Lambda\left(\!\begin{smallmatrix}
 m& i_1\\i_3& i_3\end{smallmatrix}\!\right) B_{i_2}^\dag\Big]|v\ran.
\ee
When inserted into Eq.~(\ref{app:eq:ZNtilde0}), these two terms contribute equally through a relabeling of $(i_1,i_2)$. So, the term in $D_{i_3 i_3}$ gives a contribution to $\tilde Z_3^{(0)}$ given by
\be
-2\frac{1}{3!}\sum e^{-\beta(E_{i_1}+E_{i_2}+E_{i_3})}\Lambda\left(\!\begin{smallmatrix}
m& i_2\\i_3& i_3 \end{smallmatrix}\!\right)\lan v| B_{i_1}B_{i_2} B_m^\dag  B_{i_1}^\dag |v\ran.
\ee
This term is shown in Fig.~\ref{fig:15}(b). The scalar product in the above equation gives two delta terms, namely $\delta_{ m i_2}$ and $\delta_{ m i_1}$, plus one exchange term in $\Lambda\left(\!\begin{smallmatrix}
i_2& m\\i_1& i_1 \end{smallmatrix}\!\right)$ that we can neglect if we only want the first-order correction to $\tilde Z_3^{(0)}$. The two delta terms shown in Figs.~\ref{fig:15}(c) and (d) give $-2(1/3!)L(\beta,\beta)\tilde Z_1^{(0)}$ and $-2(1/3!)L(2\beta,\beta)$ respectively.\

Finally, the term in $B_{i_3}^\dag B_{i_3}$ of Eq.~(\ref{eq:app_sp3}) is calculated by pushing $B_{i_3}$ to the right according to Eq.~(\ref{eq:commut_BB}),
\be
\lan v| B_{i_1}B_{i_2}B_{i_3}^\dag(\delta_{i_3 i_2}-D_{i_3 i_2}+B_{i_2}^\dag B_{i_3}) B_{i_1}^\dag|v\ran.
\ee
$B_{i_2}^\dag B_{i_3} B_{i_1}^\dag|v\ran=\delta_{i_3i_1}B_{i_2}^\dag |v\ran$ leads to a contribution similar to the term in $\delta_{i_3 i_2}$ through a relabeling of the $(i_1,i_2)$ indices, while $D_{i_3 i_2}B_{i_1}^\dag|v\ran$ is calculated using Eq.~(\ref{eq:commut_DB}). So, the term in $B_{i_3}^\dag B_{i_3}$ yields two terms given by
\bea
&&\frac{1}{3!}\sum e^{-\beta(E_{i_1}+E_{i_2}+E_{i_3})}\Big[2\delta_{i_3 i_2}\lan v| B_{i_1}B_{i_2} B_{i_3}^\dag  B_{i_1}^\dag |v\ran\nn\\
&&{}-\sum_m\lan v| B_{i_1}B_{i_2} B_m^\dag  B_{i_3}^\dag|v\ran \Lambda\left(\!\begin{smallmatrix}
m& i_1\\i_3& i_2 \end{smallmatrix}\!\right)\Big].\label{app:eq:tildeZ3eq}
\eea
\begin{figure}[t]
\centering
   \includegraphics[trim=5cm 2.5cm 5cm 3.4cm,clip,width=2.8in] {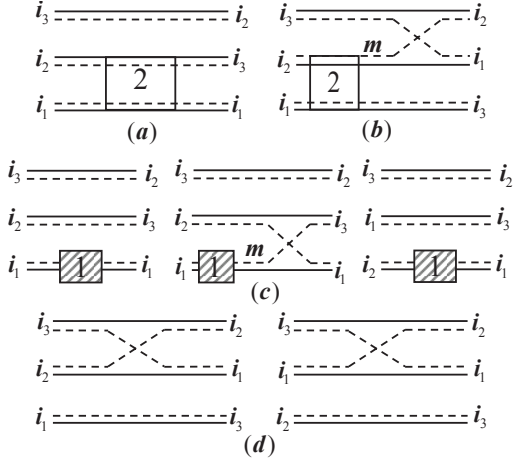}
\caption{\small Diagrams (a,b) representing the two terms in Eq.~(\ref{app:eq:tildeZ3eq}) for the three-coboson scalar product. Diagrams in (c) follow from (a), while diagrams in (d) follow from (b). }
\label{fig:16}
\end{figure}

The scalar product in the first term of the above equation, shown in Fig.~\ref{fig:16}(a), is calculated by replacing $B_{i_2} B_{i_3}^\dag$ with $\delta_{i_2 i_3}-D_{i_2 i_3}+B_{i_3}^\dag B_{i_2}$ according to Eq.~(\ref{eq:commut_BB}). Since $D_{i_2 i_3}B_{i_1}^\dag |v\ran $ gives $ \sum_m B_m^\dag|v\ran \Lambda\left(\!\begin{smallmatrix}m& i_1\\i_2& i_3 \end{smallmatrix}\!\right)$, these three terms shown in Fig.~\ref{fig:16}(c) ultimately yield a contribution to $\tilde Z_3^{(0)}$ given by
\be
\frac{1}{3!}2\Big[z(2\beta)\tilde Z_1^{(0)} -L(2\beta,\beta)+z(3\beta)\Big].
\ee
In the second term of Eq.~(\ref{app:eq:tildeZ3eq}), shown in Fig. \ref{fig:16}(b), we just have to replace the scalar product by $\delta_{i_2 m}\delta_{i_1 i_3}+\delta_{i_1 m}\delta_{i_2 i_3}$ if we want this term at first order in fermion exchange only. These two terms, shown in Fig.~\ref{fig:16}(d), yield $-2(1/3!) L(2\beta,\beta)$.

Collecting all the above terms, we end with
\bea
\tilde Z_3^{(0)}&=&\frac{1}{3}\Big[z(\beta)\tilde Z_2^{(0)}+(z(2\beta)-L(\beta,\beta))\tilde Z_1^{(0)}\nn\\
&&+(z(3\beta)-3L(2\beta,\beta))\Big],
\eea
in agreement with Eqs.~(\ref{eq:tilderrN},\ref{eq:tildezns}).
	
\renewcommand{\theequation}{\mbox{E.\arabic{equation}}} 
\setcounter{equation}{0} %
\section{ Calculation of $\tilde Z_3^{(1)}$\label{app:sec5}}
We here calculate the partition function at first order in interaction scattering, $\tilde Z_N^{(1)}$, given in Eq.~(\ref{eq:ZNtilde1}) for three cobosons, namely
\bea
\tilde Z_3^{(1)}&=&- \frac{\beta}{4}\sum e^{-\beta (E_{i_1}+E_{i_2}+ E_{i_3})}\label{app:Z3tilde1}\\
&& \times \Big[\lan v|B_{i_1}B_{i_2}  B_{i_3} B_m^\dag B_n^\dag B^\dag_{i_1} |v\ran\xi\left(\!\begin{smallmatrix}
n&i_3\\ m& i_2\end{smallmatrix}\!\right)+c.c.\Big].\nn
\eea
The above scalar product is calculated by first replacing $B_{i_3} B_{i_1}^\dag$ with $\delta_{i_3 i_1}-D_{i_3 i_1}+B_{i_1}^\dag B_{i_3}$ according to Eq.~(\ref{eq:commut_BB}).\\
(i) The $\delta_{i_3 i_1}$ term leads to a contribution to $\tilde Z_3^{(1)}$ given by
\be
- \frac{\beta}{4}\sum e^{-\beta (E_{i_2}+ 2E_{i_3})}\Big[\lan v|B_{i_3}B_{i_2}  B_m^\dag B_n^\dag  |v\ran\xi\left(\!\begin{smallmatrix}
n&i_3\\ m& i_2\end{smallmatrix}\!\right)+c.c.\Big].\label{app5:i1}
\ee
As $\lan v|B_{i_3}B_{i_2}  B_m^\dag B_n^\dag  |v\ran=\delta_{i_3m}\delta_{i_2n}+\delta_{i_3n}\delta_{i_2m}-\Lambda\left(\!\begin{smallmatrix}
i_3& m\\ i_2& n\end{smallmatrix}\!\right)$, we ultimately get this contribution to $\tilde Z_3^{(1)}$ as
\be
-2 \frac{\beta}{4}\hat \xi(\beta,2\beta)
\ee
with $\hat \xi(n_1\beta,n_2\beta)$ defined in Eqs.~(\ref{def:hatxin1n2},\ref{def:hatxii1i2}), the factor of 2 coming from the $c.c.$ part.\\
(ii) The term in $D_{i_3 i_1}$, inserted into Eq.~(\ref{app:Z3tilde1}), leads to
\bea
\lefteqn{\frac{\beta }{4}\sum e^{-\beta (E_{i_1}+E_{i_2}+ E_{i_3})}}\label{app5:D31}\\
&&\times\Big[\lan v|B_{i_1}B_{i_2}  D_{i_3i_1} B_m^\dag B_n^\dag  |v\ran\xi\left(\!\begin{smallmatrix}
n&i_3\\ m& i_2\end{smallmatrix}\!\right)+c.c.\Big].\nn
\eea
Using Eq.~(\ref{eq:commut_DB}), we get $D_{i_3i_1} B_m^\dag B_n^\dag  |v\ran$ as
\be
\sum_p \Lambda\left(\!\begin{smallmatrix}
p&m\\ i_3& i_1\end{smallmatrix}\!\right)B^\dag_pB^\dag_n |v\ran+(m\longleftrightarrow n).
\ee
So, Eq.~(\ref{app5:D31}) gives
\bea
\lefteqn{2\frac{\beta }{4}\sum e^{-\beta (E_{i_1}+E_{i_2}+ E_{i_3})}}\\
&&\times\Big[\lan v|B_{i_1}B_{i_2} B_p^\dag B_n^\dag  |v\ran \sum_m \Lambda\left(\!\begin{smallmatrix}
p&m\\ i_3& i_1\end{smallmatrix}\!\right)\xi\left(\!\begin{smallmatrix}
n&i_3\\ m& i_2\end{smallmatrix}\!\right)+c.c.\Big].\nn
\eea
The sum over $m$ corresponds to a scattering represented by a diagram similar to the one of Fig.~\ref{fig:10}(c). As it involves three cobosons, this term leads to a contribution to $\tilde Z_3^{(1)}$ of the order of $(a_X^3/L^3)^2$ which can be neglected in a first-order calculation.\\
(iii) The term in $B_{i_1}^\dag B_{i_3}$ leads to a contribution to $\tilde Z_3^{(1)}$ given by
\bea
\lefteqn{- \frac{\beta}{4}\sum e^{-\beta (E_{i_1}+E_{i_2}+ E_{i_3})}}\\
&& \times\Big[\lan v|B_{i_1}B_{i_2} B^\dag_{i_1} B_{i_3} B_m^\dag B_n^\dag  |v\ran\xi\left(\!\begin{smallmatrix}
n&i_3\\ m& i_2\end{smallmatrix}\!\right)+c.c.\Big].\nn
\eea
To get it, we replace $B_{i_2} B^\dag_{i_1}$ by $ \delta_{i_1i_2}-D_{i_2i_1}+B^\dag_{i_1}B_{i_2}$: The term in $\delta_{i_1i_2}$ is equivalent to the one of Eq.~(\ref{app5:i1}) if we interchange $i_2$ and $i_3$; so, it gives a contribution equal to
\be
-2 \frac{\beta}{4}\hat \xi(2\beta,\beta).
\ee
The term in $D_{i_2i_1}$ leads to
\bea
\lefteqn{\frac{\beta}{4}\sum e^{-\beta (E_{i_1}+E_{i_2}+ E_{i_3})}}\label{app5:D21}\\
&&\times\Big[\lan v|B_{i_1}D_{i_2i_1} B_{i_3} B_m^\dag B_n^\dag  |v\ran\xi\left(\!\begin{smallmatrix}
n&i_3\\ m& i_2\end{smallmatrix}\!\right)+c.c.\Big].\nn
\eea
Since the above scalar product already contains one fermion exchange associated with $D_{i_2i_1}$, we can reduce $B_{i_3} B_m^\dag B_n^\dag  |v\ran $ to $\delta_{i_3m}  B_n^\dag  |v\ran+\delta_{i_3n} B_m^\dag  |v\ran$ at lowest order in $a_X^3/L^3$. When inserted into Eq.~(\ref{app5:D21}), we get
 \bea
2\frac{\beta}{4}\sum e^{-\beta (E_{i_1}+E_{i_2}+ E_{i_3})}\Big[\lan v|B_{i_1}D_{i_2i_1}  B_n^\dag  |v\ran\xi\left(\!\begin{smallmatrix}
n&i_3\\ i_3& i_2\end{smallmatrix}\!\right)+c.c.\Big].\nn
\eea
As $\lan v|B_{i_1}D_{i_2i_1}  B_n^\dag  |v\ran$ reduces to $\Lambda\left(\!\begin{smallmatrix}
i_1&n\\ i_2& i_1\end{smallmatrix}\!\right)$, the term in $D_{i_2i_1}$ leads to a scattering involving three cobosons; so, it gives a contribution of the order $(a_X^3/L^3)^2$ which can be neglected at lowest order. The term in $B^\dag_{i_1}B_{i_2}$ gives
\bea
\lefteqn{-\frac{\beta}{4}\sum e^{-\beta (E_{i_1}+E_{i_2}+ E_{i_3})}}\label{app5:B1B2}\\
&&\times\Big[\lan v|B_{i_1}B^\dag_{i_1}B_{i_2} B_{i_3} B_m^\dag B_n^\dag  |v\ran\xi\left(\!\begin{smallmatrix}
n&i_3\\ m& i_2\end{smallmatrix}\!\right)+c.c.\Big].\nn
\eea
As $\lan v|B_{i_1}B^\dag_{i_1}=\lan v|\delta_{i_1i_1}$, the above contribution reduces to
\bea
&&\hspace{-0.5cm}z(\beta)\frac{-\beta}{4}\sum e^{-\beta (E_{i_2}+ E_{i_3})}\big[ \lan v|B_{i_2} B_{i_3} B_m^\dag B_n^\dag  |v\ran\xi\left(\begin{smallmatrix}
n&i_3\\ m& i_2\end{smallmatrix}\!\right)+ c.c. \big] \nn\\
&&= z(\beta)\tilde Z_2^{(1)}=-\frac{\beta}{2}\hat \xi(\beta,\beta)\tilde Z_1^{(0)}.
\eea

All these terms combine to yield, with $\hat \xi(n\beta)$ defined in Eq.~(\ref{def:hatxinbeta}),
\be
\tilde Z_3^{(1)}=-\frac{\beta}{2}\Big[\hat \xi(2\beta)\tilde Z_1^{(0)}+\hat \xi(3\beta)\Big],
\ee
in agreement with Eq.~(\ref{def:tildeZN1}).

\end{document}